\documentstyle[11pt,epsf]{article}
\newcommand{\rf}[1]{(\ref{#1})}
\newcommand{\beq}{\begin{equation}}
\newcommand{\eeq}{\end{equation}}
\newcommand{\bdm}{\begin{displaymath}}
\newcommand{\edm}{\end{displaymath}}
\newcommand{\bea}{\begin{eqnarray}}
\newcommand{\eea}{\end{eqnarray}}

\newcommand{\pa}{\partial}

\newcommand{\br}{{\bf r}}
\newcommand{\bu}{\bf u}
\newlength{\bredde}
\def\slash#1{\settowidth{\bredde}{$#1$}\ifmmode\,\raisebox{.15ex}{/}
\hspace*{-\bredde} #1\else$\,\raisebox{.15ex}{/}\hspace*{-\bredde} #1$\fi}

\def\gsim{\raise.3ex\hbox{$>$\kern-.75em\lower1ex\hbox{$\sim$}}}
\def\lsim{\raise.3ex\hbox{$<$\kern-.75em\lower1ex\hbox{$\sim$}}}

\def\vek#1{{\bf #1}}
\begin{document}
\topmargin -1.4cm
\oddsidemargin -0.8cm
\evensidemargin -0.8cm
\topskip 9mm   
\headsep 9pt
\title{Turbulent Binary Fluids: A Shell Model Study. }
\author{\sc{Mogens H. Jensen$^1$ and Poul Olesen$^2$ }\\{\sl The Niels Bohr
Institute,
University of Copenhagen}\\{\sl Blegdamsvej 17, DK-2100 Copenhagen, Denmark}}
\maketitle
\vfill
\begin{abstract}   
We introduce a shell (``GOY'') model for turbulent binary fluids. The variation
in the concentration between the two fluids acts as an active scalar leading
to a redefined conservation law for the energy, which is
incorporated into the model together with a conservation law for the
scalar. The model is studied numerically at very high
values of the Prandtl and Reynolds numbers and we investigate the properties
close to the critical point of the miscibility
gap where the diffusivity vanishes. A peak develops in the spectrum of the
scalar, showing that a strongly turbulent flow leads to an  increase in
the mixing time. The peak is, however, not very pronounced.
The mixing time diverges with the Prandtl number as a power
law with an exponent $\approx$ 0.9. The continuum limit of the shell equations
leads to a set of equations which can be solved by a scaling ansatz,
consistent with an exact scaling of the Navier-Stokes equations in the
inertial range. In this case a weak peak also persists for a certain time in
the spectrum of the scalar. Exact analytic solutions of the continuous shell
equations are derived in the inertial range. Starting with fluids at rest, from
an initial variation of the concentration difference, one can provoke a
``spontaneous'' generation of a velocity field, analogous to MHD in the
early universe.
\end{abstract}
\vfill
\begin{flushleft}
cond-mat/9610047
\end{flushleft}
\footnoterule
{\small    $^1$E-mail: mhjensen@nbi.dk,  $^2$E-mail: polesen@nbi.dk}
\thispagestyle{empty}
\newpage
\setcounter{page}{1}

\section{Introduction: Turbulent fluid mixtures.}

Binary fluid mixtures provide a beautiful example of physical systems where
it is possible to study the behavior in the limit of exceedingly
large values of the Prandtl number. This is the case for miscible
binary mixtures just above the consolute temperature $T_c$, at which
the diffusivity vanishes as a power law \cite{Henry}
\beq
D(T) \simeq D_0 \left( {{T-T_c} \over T_c} \right)^\mu
\label{m1}
\eeq
 with an exponent in the range
\beq
\mu \sim 0.63 - 0.74 ~~.
\label{m2}
\eeq
Since $D(T) \to 0$ at the critical point, the corresponding
Prandtl number $Pr = \nu/D \to \infty$ and values as high as $10^6$
can be obtained experimentally \cite{Henry}. Goldburg and coworkers
studied turbulent binary mixtures experimentally using light scattering 
techniques \cite{Walter1,Walter2,Walter3,Walter4}. One can measure both
the variation of the mixing times and the growth of domains using these
techniques and we return to a discussion of the experimental results later.

Our motivation for the present work is that we are able to formulate
the theoretical equations behind binary fluid mixtures in terms of shell 
models where the known invariant quantities are conserved. In this way one
can investigate the mixture at much higher values of the Reynolds and
Prandtl numbers, than is possible by standard numerical simulations.
For instance, one can then study the scaling behavior of the mixing time
at very high values of the Prandtl number, a regime which might
be experimentally accessible. Also, the shell model is known to include 
intermittency effects, which have not been treated in previous theoretical
works on binary fluid mixtures.

With the two fluids labelled $A$ and $B$, respectively, a scalar field
is defined as \cite{Siggia,Hohenberg}
\beq
\psi({\br} ,t) = (\rho_A ({\br} ,t) - \rho_B ({\br} ,t))/\rho_0
\label{m3}
\eeq
where $\rho_A({\br},t)$ and $\rho_B({\br},t)$ are the mass densities of the
two fluids and $\rho_0$ is the mean mass density. In a phase plane
determined by the temperature $T$ versus the average of the scalar
$<\psi>$, there exists
a ``miscibility-gap'' separating the miscible phase from the immiscible.
Along this separating curve, the effective diffusivity $D$ vanishes, because
it separates a regime where the effective diffusivity is positive
(i.e the miscible case) from a regime where it is negative (the 
immiscible case). Close
to the 50-50 \% concentrations of the two fluids,
one finds in equilibrium 
\beq
< \psi ({\br},t) > ~= 0~~.
\label{m4}
\eeq
For this case, as the critical point is approached from the miscible
phase, the scalar is supposed to be ``active'' and influence the
velocity equation of the Navier-Stokes equations quite substantially.
The corresponding equations of motion were derived almost two decades ago
by Siggia, Halperin and Hohenberg \cite{Siggia,Hohenberg} and later
on elaborated quite a lot by Ruiz and Nelson \cite{Ruiz1,Ruiz2}
\beq
{{\pa \psi} \over {\pa t}} + ({\bu} \cdot {\bf \nabla}) \psi = D \nabla^2 \psi
\label{m5}
\eeq
\beq
{{\pa {\bu}} \over {\pa t}} + ({\bu} \cdot {\bf \nabla}){\bu} =
- {1 \over {\rho_0}} {\bf \nabla} p' - \alpha {\bf \nabla} \psi 
\nabla^2 \psi + \nu \nabla^2 {\bu} + {\bf f}
\label{m6}
\eeq
\beq
{\bf \nabla} \cdot {\bu} =0
\label{p1}
\eeq
Here $\nu$ is the kinematic viscosity; ${\bf f}$ is the forcing and
several terms involving $\psi$ have been incorporated into an effective
pressure $p'$ \cite{Siggia,Hohenberg}. 
The term with coefficient $\alpha$ represents
the ``active'' part of the scalar. This term acts like a force of
the form $\mu_{AB} {\bf \nabla} \psi$ where $\mu_{AB} = -\alpha 
\nabla^2 \psi$ plays the role of a local chemical potential difference
between the $A$ and $B$ component of the mixture \cite{Ruiz1}-\cite{Ruiz2}.
The coefficient $\alpha$ has the dimensions of the square of
a transport coefficient
and has been estimated to be of the order $\alpha \sim \nu^2$ in reference
\cite{Ruiz1}.

The equations of motion \rf{m5} and \rf{m6} allow two quadratic invariants
in the absence of diffusivity, viscosity and forcing, i.e. in the limit
$D = \nu = 0$, ${\bf f} = {\bf 0}$. The first is the squared integral
of the concentration fluctuations
\beq
C_{tot} = {1 \over 2} \int d \br (\psi(\br,t))^2
\label{m7}
\eeq
and the second is the total energy with a term relating to the
active influence of the scalar
\beq
E_{tot} = {1 \over 2} \int d {\br} ( \mid {\bu} ({\br},t) \mid^2
+ \alpha \mid {\bf \nabla} \psi ({\br},t) \mid^2 )~~.
\label{m8}
\eeq
In the case of a passive scalar, i.e. when $\alpha = 0$, one expects
the energy spectrum $E(k)$ and the spectrum of the scalar $C(k)$ to have the
usual behavior
\beq
E(k) \sim k^{-5/3 - \delta}  ~~~~,~~~~ C(k) \sim k^{-5/3 - \gamma}
\label{m9}
\eeq
where -5/3 is the Kolmogorov exponent \cite{Kol} (and 
Obukhov-Corrsin \cite{Obukhov} exponent for the scalar) and $\delta$ and
$\gamma$ are intermittency corrections in the 
two cases, respectively \cite{mogens}.
Ruiz and Nelson \cite{Ruiz1,Ruiz2} also discuss the possibility of internal
wave-like excitations, in the case of large values of $\alpha$, similar
to linear wave excitations in MHD, which may change the spectrum to
different scaling behavior as predicted by Iroshnikov and
Kraichnan \cite{Iroshnikov,Kraichnan}.
We do not discuss this phenomenon here but reserve it for a forthcoming
publication.

An interesting feature of the equations of motion \rf{m5} and \rf{m6}
should be noticed: If the initial velocity field $\bf u$ vanishes for $t$=0,
then
if
\beq
{\bf \nabla} \psi \not =0 ~{\rm and}~\nabla^2 \psi \not =0,
 ~{\rm and/or}~{\bf \nabla}p'\not=0~
~{\rm for}~t=0,
\label{mnew}
\eeq
it follows from Eq. \rf{m6} that a finite velocity field will appear. For 
small times it is given by
\beq
{\bf u}({\bf x},t)=-\left(\frac{1}{\rho_0}\nabla p'({\bf x},0)+\alpha 
{\bf \nabla}\psi({\bf x},0)\nabla^2\psi({\bf x},0)\right)t+O(t^2).
\label{mny}
\eeq 
Here the pressure term should be such as to respect eq. \rf{p1}. Eq. \rf{mny}
means that if initially the liquids are at rest, and experimental
initial conditions respecting \rf{mnew} are established, then one should see
that the liquids start to move ``spontaneously''. This is a very clean effect
of a non-vanishing ``transport coefficient'' $\alpha$, and it works
irrespectively of the magnitude of the diffusion coefficient $D$. This will
be further discussed in section 7.

The paper is organized as follows. In section 2 we derive the shell model
for turbulent binary fluid mixtures and discuss the corresponding conservation
law, the value of the 
coupling constants, etc. In Section 3 the numerical results
obtained from integrating the model are presented. In particular we discuss 
the appearance of a peak in the spectrum of the scalar. Section 4 contains
the theoretical predictions of Ruiz and Nelson for the mixing times and
the corresponding results from the shell model. In section 5 we present the
continuum version of the ``GOY'' model and exact analytic solutions in
the inertial range, based on a scaling ansatz. In section 6 the corresponding
continuum equations for
the binary mixture model are derived and in section 7 the numerical results
from integrating these equations are presented together with a comparison
with the results from the discrete equations. Finally, section 8 offers
concluding remarks.

\section{A shell model for binary mixtures.}

Since the binary mixtures are particularly interesting to investigate in the
critical regime where $Pr \to \infty$ and as we are concerned with the
case of a strongly turbulent mixture (large values of $Re$),
it is our goal to formulate an approximate scheme for Eqs. \rf{m5} and \rf{m6} 
in which this limit is accessible. Shell models in Fourier space fulfill
these requirements. They have been introduced by Obukhov \cite{Obukhov1}, 
Gledzer \cite{gledzer}, Desnyansky and Novikov \cite{DN}. The key idea is to
mimic the Navier-Stokes equations by a dynamical system with $N$ variables
$u_1, u_2,...., u_N$, each of which representing the typical magnitude of
the velocity field on a certain length scale. The Fourier space
is divided in $N$ shells and each shell consists of the set of
wavevectors ${\bf k}$ such that $k_0 r^n < ~\mid {\bf k} \mid~ < k_0 r^{n+1}$.
The variable $u_n$ is the velocity difference over a length $\sim k_n^{-1}$
so that there is only one degree of freedom per shell \cite{mogens}. 
Also models with a large number of degrees of freedom have been introduced 
and analysed \cite{Grossmann,Aurell}. The most studied model is the ``GOY''
model introduced by Ohkitani and Yamada \cite{Ok-Ya} which was found to
be intermittent by Jensen, Paladin and Vulpiani \cite{JPV} and 
studied extensively in many other contexts 
\cite{kadf,benzi,bif1,brandenburgetal,parisi,bif}. This model uses
af complex set of variables and has the same type of quadratic non-linearities
and the same symmetries as the 3d Navier-Stokes equations. We shall
apply the same approach in this paper by expanding the ``GOY'' model 
to include the term of the active scalar and at the same time conserve
the two different quadratic non-linearities \rf{m7} and \rf{m8}. 

Firstly, we write down the shell model for the 
scalar equation \rf{m5} \cite{Ruiz1,JPV1}. 
Using a complex field $\psi_n$ associated to shell $n$, the equation becomes
\bea
({d\over dt}+ D k_n^2 ) \ \psi_n \  &=& i \ [
 e_n~k_n~  (u^*_{n-1} \psi^*_{n+1} \ - \ u^*_{n+1} \psi^*_{n-1} ) 
+ g_n~k_{n-1}~  (u^*_{n-2} \psi^*_{n-1} \ + \ u^*_{n-1} \psi^*_{n-2} ) 
\nonumber \\ &+& 
   h_n~k_{n+1}~ (u^*_{n+1} \psi^*_{n+2} \ + \ u^*_{n+2} \psi^*_{n+1} )  \ ] 
\label{m10}
\eea
where $D$ is the molecular diffusion. The physical time scale of the model is 
determined by the the constant
$k_1 = r k_0$, which represents the inverse scale of the largest eddy, and
the related velocity $|u_1|$. This time scale is therefore the
corresponding eddy-turn-over time $1/(k_1 |u_1|)$. In the following,
``time unit'' thus means this time scale.
                                                 
The  coefficients of the advective terms  follow
from demanding  the conservation  of $\sum_n |\psi_n|^2$ 
when the diffusivity is vanishing $D = 0$. A possible choice is:
\beq
e_n={1\over r} \qquad g_n=-{1\over r} \qquad
h_n={1\over r}
\label{m11}
\eeq
with
\beq
e_1=e_N=g_1=g_2=h_{N-1}=h_N=0
\eeq
The shell model equations for \rf{m6} will consist of two contributions; the 
first
part is the usual ``GOY'' shell model \cite{Ok-Ya} (with coefficients
$a_n , b_n , c_n$) and the second part is the shell expression for the
``active'' term in the velocity equation \rf{m6}
\bea
({d\over dt}+\nu k_n^2 ) \ u_n \ &=& 
 i~k_n~ [ a_n \,  u^*_{n+1} u^*_{n+2} \, + \, {b_n} u^*_{n-1} u^*_{n+1} \, + \,
  {c_n }  u^*_{n-1} u^*_{n-2} ] \nonumber \\
&+& i ~ \alpha ~ k_n^3 ~ [r_n  \psi^*_{n+1} \psi^*_{n+2} ~+~ s_n \psi^*_{n-1} 
\psi^*_{n+1}
~+~ t_n \psi^*_{n-1} \psi^*_{n-2} ] \ + \ f \delta_{n,4}
\label{m12}
\eea
with $n=1,\cdots N$,  $k_n=r^n \, k_0$ and  boundary conditions
\beq
b_1=b_N=c_1=c_2=a_{N-1}=a_N=s_1=s_N=t_1=t_2=r_{N-1}=r_N=0
\eeq
In order to ensure the conservation of the quadratic quantity 
\beq
E_{tot} = \sum_n \left(\mid u_n \mid^2 + \alpha k_n^2 \mid \psi_n \mid^2\right)
\label{m13}
\eeq
in the limit without viscosity, diffusivity and forcing, $\nu = D = f = 0$, one
multiplies Eq.\rf{m12} by $u_n$  and multiplies Eq.\rf{m10} 
by $\psi_n$ and then balances the terms. The non-linear
terms in $u_n$ lead to the usual constraints of 
the ``GOY'' model \cite{mogens}
\beq
a_n =1 ~,~ b_n = - {\delta \over r} ~,~ c_n = - {{1 - \delta} \over {r^2}}~.
\label{m14}
\eeq
For the second part of Eq.\rf{m12} one balances the terms by the corresponding
terms in the scalar equation. With the choice of the parameters \rf{m11}
we then obtain the following conditions for the coefficients 
\beq
r_n ~=~r^4 - r^2 ~,~ s_n ~= ~r - r^{-3} ~,~
t_n ~= ~r^{-4} - r^{-6}~~.
\label{m15}
\eeq
One observes that when the coupling constants for the
scalar equation are given \rf{m11}, then the coefficients 
of the ``active'' terms are fixed. As these active terms are
proportional to $k_n^3$, we expect the effects of this term to
show up at the end of the spectrum for large values of $k_n$.

\section{Results from the shell model.}

\begin{figure}[htbp]
\mbox{{\epsfxsize=3.2in \epsfysize=3.2in \epsfbox{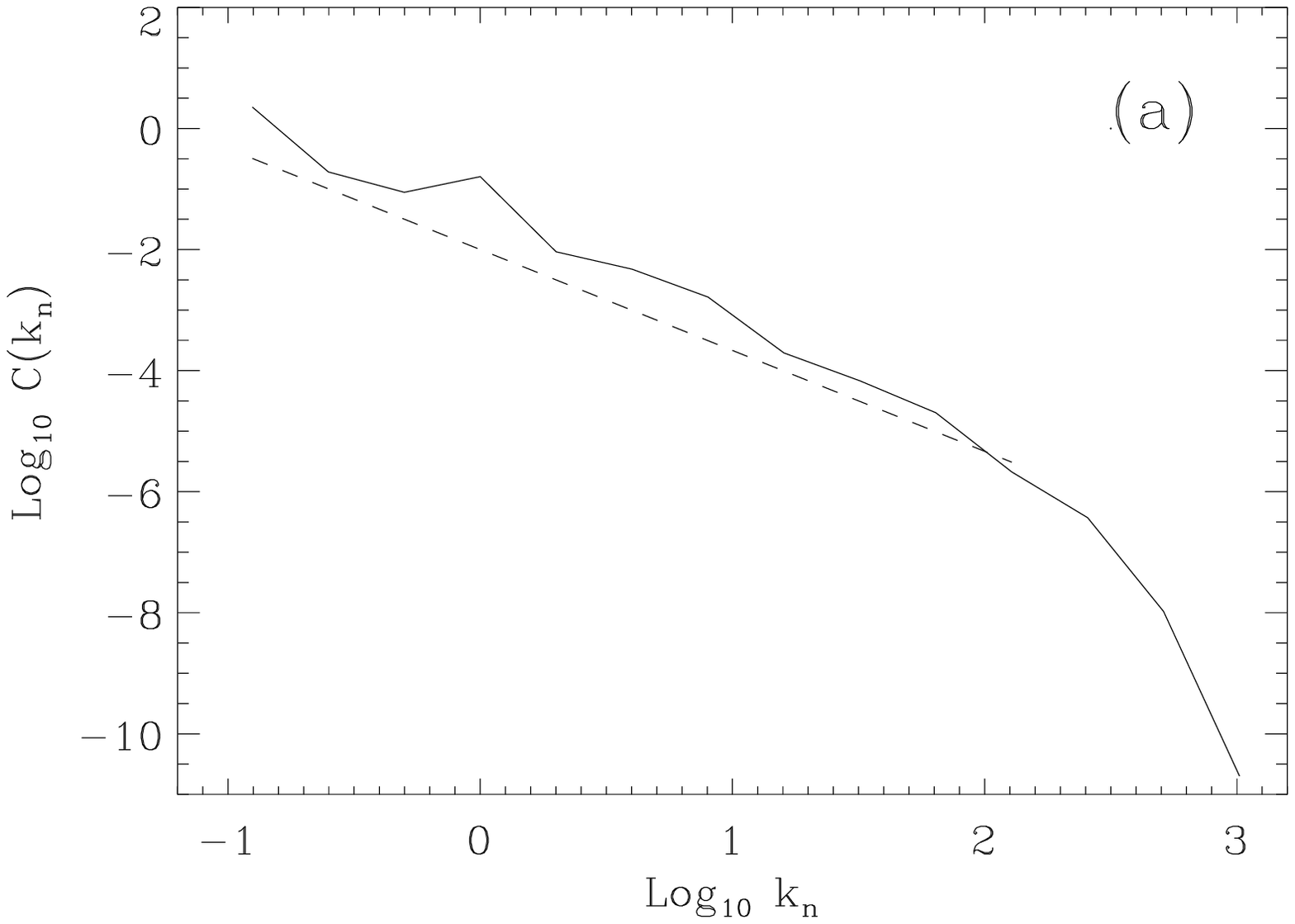}}
\vspace*{.5cm}
\hspace*{.1cm}
{\epsfxsize=3.2in \epsfysize=3.2in \epsfbox{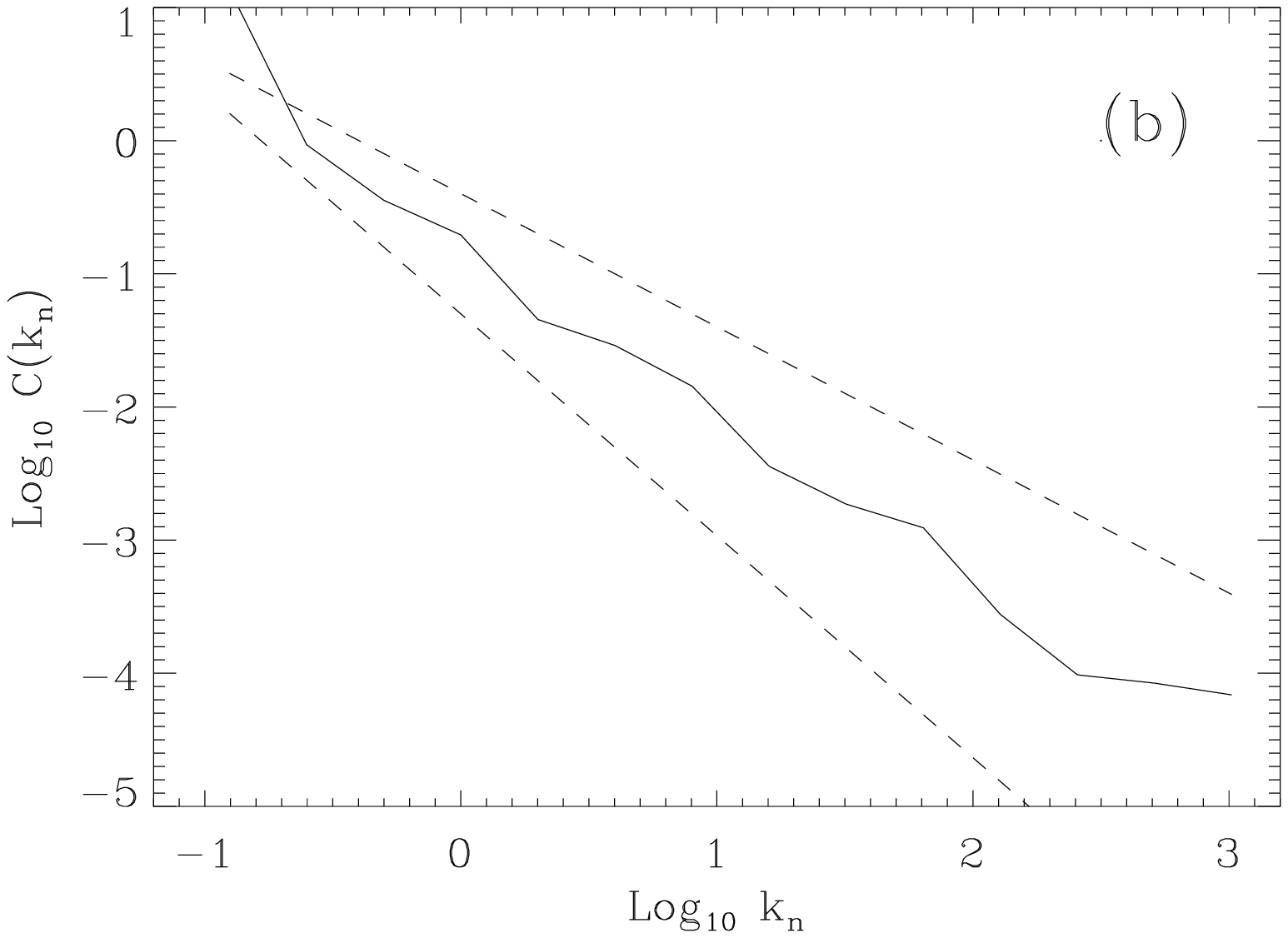}}}
\mbox{{\epsfxsize=3.2in \epsfysize=3.2in \epsfbox{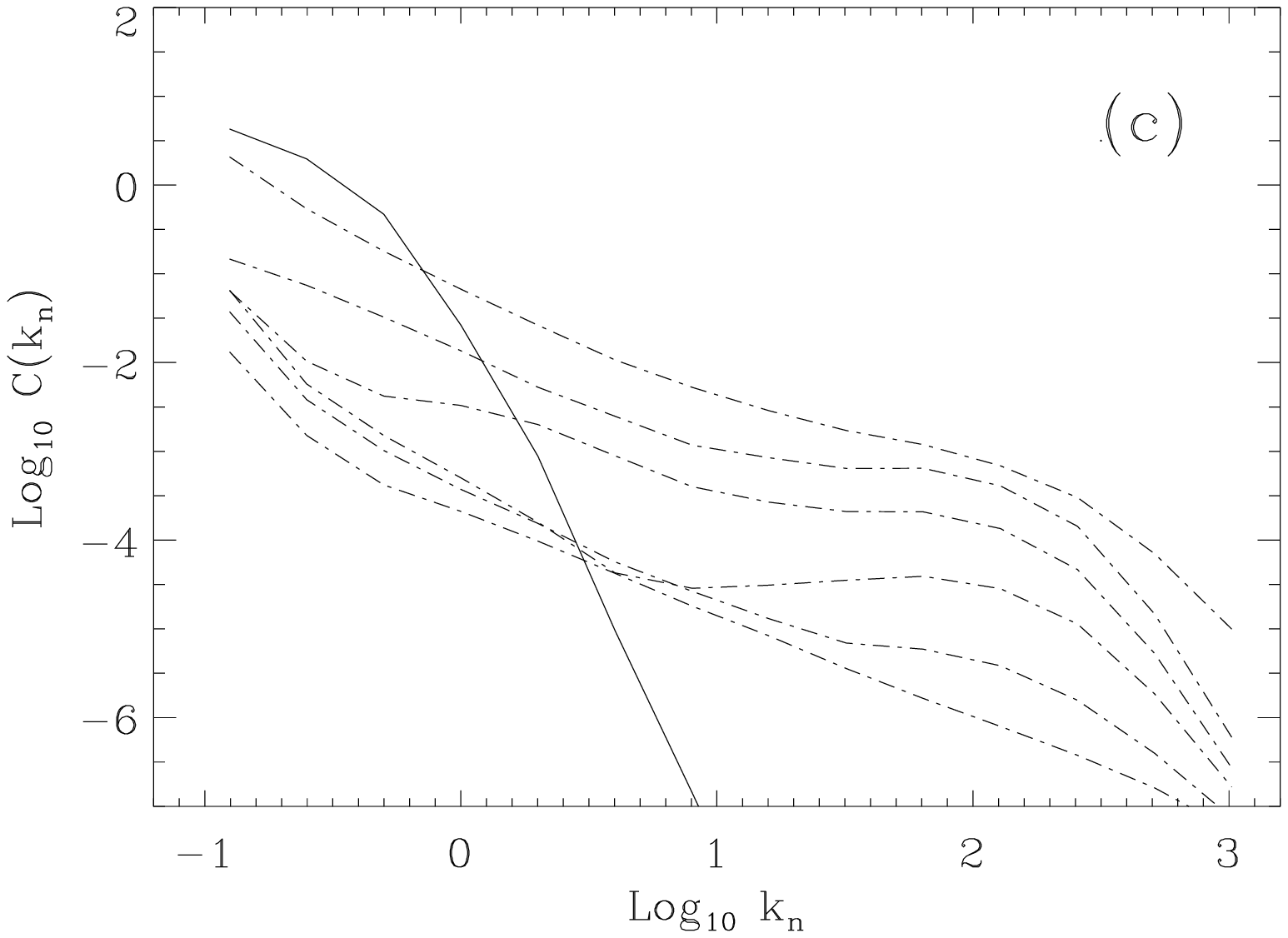}}
\vspace*{.5cm}
\hspace*{.1cm}
{\epsfxsize=3.2in \epsfysize=3.2in \epsfbox{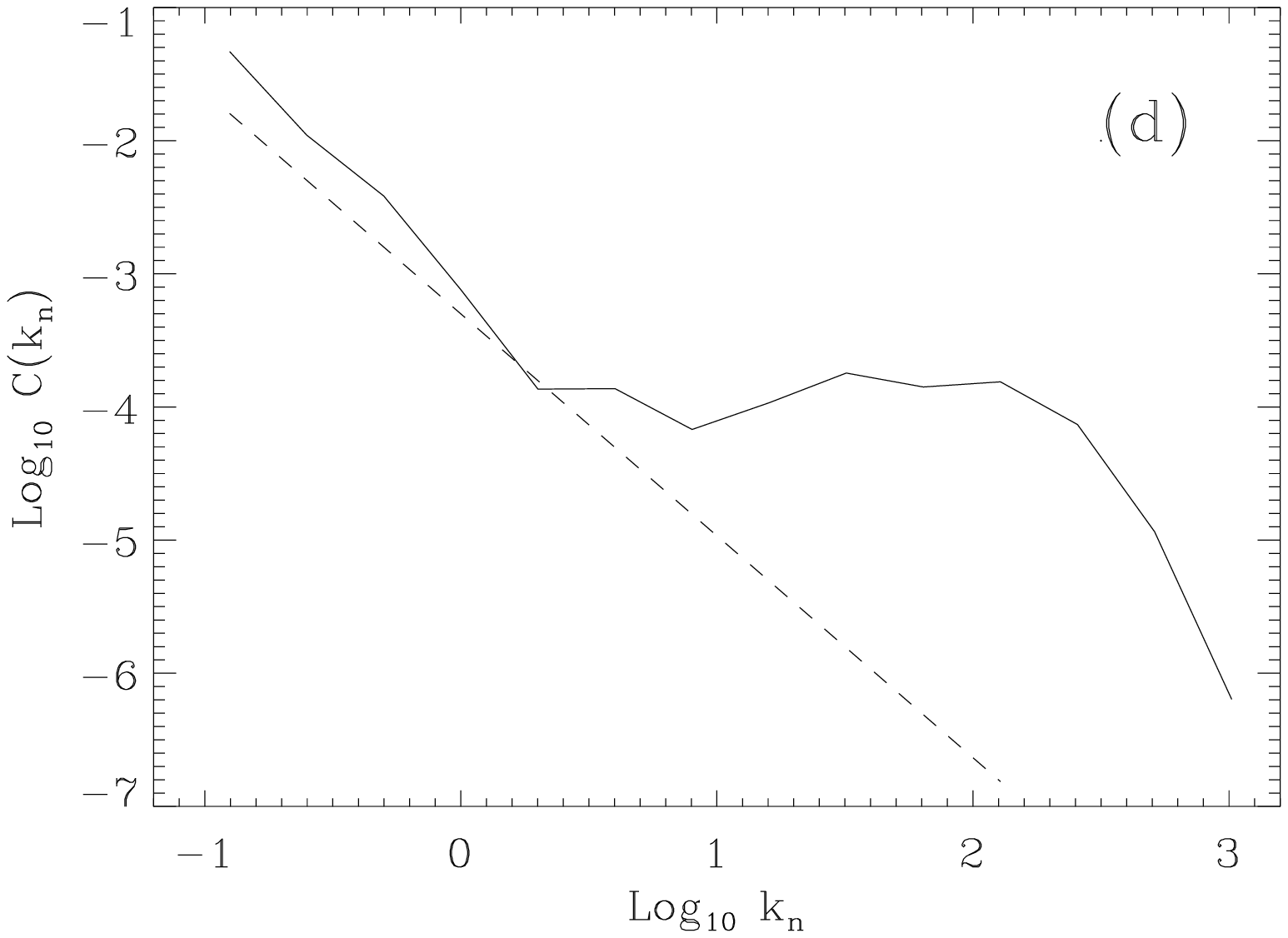}}}
\vspace*{-1cm}
\caption[xxx]{
Results from numerical integrations of the shell model \rf{m10} and \rf{m12}.
The parameters are: $N$=14, $k_0$ = 2$^{-4}$, $\nu$ = 10$^{-4}$. The 
spectrum in (a) is averaged over 4000 time units; the other 
over 200 time units. In the case (a) the spectrum
 $C(k_n)$
has $D$ =10$^{-4}$, $Pr$ =1, and $\alpha =0$. The dashed line in (a) has slope
-5/3. In the case (b) $D$=10$^{-7}$, $Pr$=10$^3$, and $\alpha$=0, and the
dashed lines have slopes -1 and -5/3, respectively. (c): $C(k_n)$ for
 $D$ = 10$^{-7}$ and $\alpha$ = 10$^{-8}$.
The full curve shows the initial condition. Time progresses from
the uppermost spectrum to the lower. The interval between the curves
are of the order $\sim$2000 time units. In (d) we show the spectrum at one
particular time. The dashed line has slope -5/3.} 
\label{fm1a}
\end{figure}

This section contains some of the numerical results obtained 
from integrations of the shell model for the binary mixture, derived above.
In the simulations we use the standard separation between
the shells, $r =2$, such that $k_n = 2^n k_0$. We apply the ``symmetric
choice'' of the ``GOY'' parameters, $\delta = 1/2$. In this
case it is known that for the ``GOY'' model alone, the second
quadratic invariant assumes the symmetry of a helicity \cite{kadf}, 
and for that case the model is strongly intermittent and gives results
in good agreement with experiments.
We study the model with $N=14$ and 19 shells, $k_0 = 2^{-4}$, and the strength 
of the forcing term in \rf{m12} is $f = 0.005 \cdot (1+i)$.
As argued by Ruiz and Nelson \cite{Ruiz1,Ruiz2} the coupling constant
of the active term $\alpha$, has the dimensions of a square of a transport 
coefficient and is in the order of magnitude
\beq
\alpha \sim \nu^2~~~~.
\label{m16}
\eeq
The spectrum of the scalar for the shell variables is defined as
\beq
C(k_n) ~=~ < \mid \psi_n \mid^2 > / k_n~~.
\label{m17}
\eeq
The brackets stand for averages over initial conditions and time.
Similarly, the energy spectrum is defined as
\beq
E(k_n) ~=~ < \mid u_n \mid^2 > / k_n ~~.
\eeq
\label{m18}
Since the scalar equations \rf{m5},\rf{m10} are not forced, then $C(k_n) \to 0$
in the long time limit. Nevertheless, it is possible 
to obtain intermediate averages
over shorter times. This is in contrast to the case of
the velocity spectrum $E(k_n)$ where
the mean exists for $t \to \infty$ as the velocity equations 
\rf{m6},\rf{m12} are forced. Firstly, we present results when the value 
of the viscosity is $\nu = 10^{-4}$,
meaning that $Re \sim 10^4$. For this value of the viscosity, a shell
model with $N$=14 shells is employed.
In order to observe the differences between 
a passive and an active scalar, we first consider the case $\alpha =0$.
Fig. \ref{fm1a}a shows the corresponding spectrum $C(k_n)$ on logarithmic
scales
in the case $Pr =1$. The spectrum follows quite closely 
the Obukhov law \cite{Obukhov}.

Next the Prandtl number is increased to 
$Pr = 10^{3}$. The spectrum, shown in Fig. \ref{fm1a}b, is changed
and scales for high value of $k_n$ according
to the Batchelor law $C(k_n) \sim k_n^{-1}$, which means that
with the increased value of the Prandtl number a viscous-convective regime is
observed, as expected \cite{Ruiz1}. 
When the scalar becomes active, i.e. $\alpha \not= 0$, by introducing 
the coupling term \rf{m16},
the spectrum changes completely and a peak develops at the upper end 
of the spectrum as indicated on the evolution series, Fig. \ref{fm1a}c.
The series is initiated in a state where $\psi_n$
is concentrated at a low $k_n$ value. This corresponds to a
large scale disturbance, for instance where the fluids $A$ and $B$ are
completely separated. After some time, a peak (or, perhaps more appropriate,
a ``shoulder'') develops at large values of $k_n$ 
as indicated in the figure
(each curve is averaged over 200 time units). After further time the 
peak disappears and the spectrum ends up in $C(k_n) \sim k_n^{-1}$.
The final spectrum in Fig. \ref{fm1a}c represents again an almost stationary
situation and does not change significantly 
during long time. Let us
note in passing that time-averaging give results similar to ensemble-
averaging, since the dynamics in the phase space of the $u_n$- and
$\psi_n$-field is strongly chaotic as determined by positive Lyapunov
exponents \cite{mogens}.

This peak was predicted by Ruiz and Nelson and also seen in numerical
simulations using Markorvian closure equations \cite{Ruiz2}.
In fact, in those simulations the peak appears much more pronounced than
compared to our results. We believe the reason is that the strongly
intermittent motion could influence the dynamics in a way that the peak
becomes less strong.
Fig. \ref{fm1a}d shows one of the spectra in the series indicating the
peak more clearly. Note, that for low $k_n$-values, the spectrum is still
close to the $- {5 \over 3}$  law as indicated by the dashed line. 
The peak indicates, that two miscible fluids close
to the critical point mix very slowly at the small scales when the
fluid is strongly turbulent. The time for which the peak persists
is strongly dependent on the Prandtl and Reynolds numbers. This time
is called the mixing time \cite{Ruiz2}.

\begin{figure}[htb]
\hbox{{\epsfxsize=3.2in \epsfysize=3.2in \epsfbox{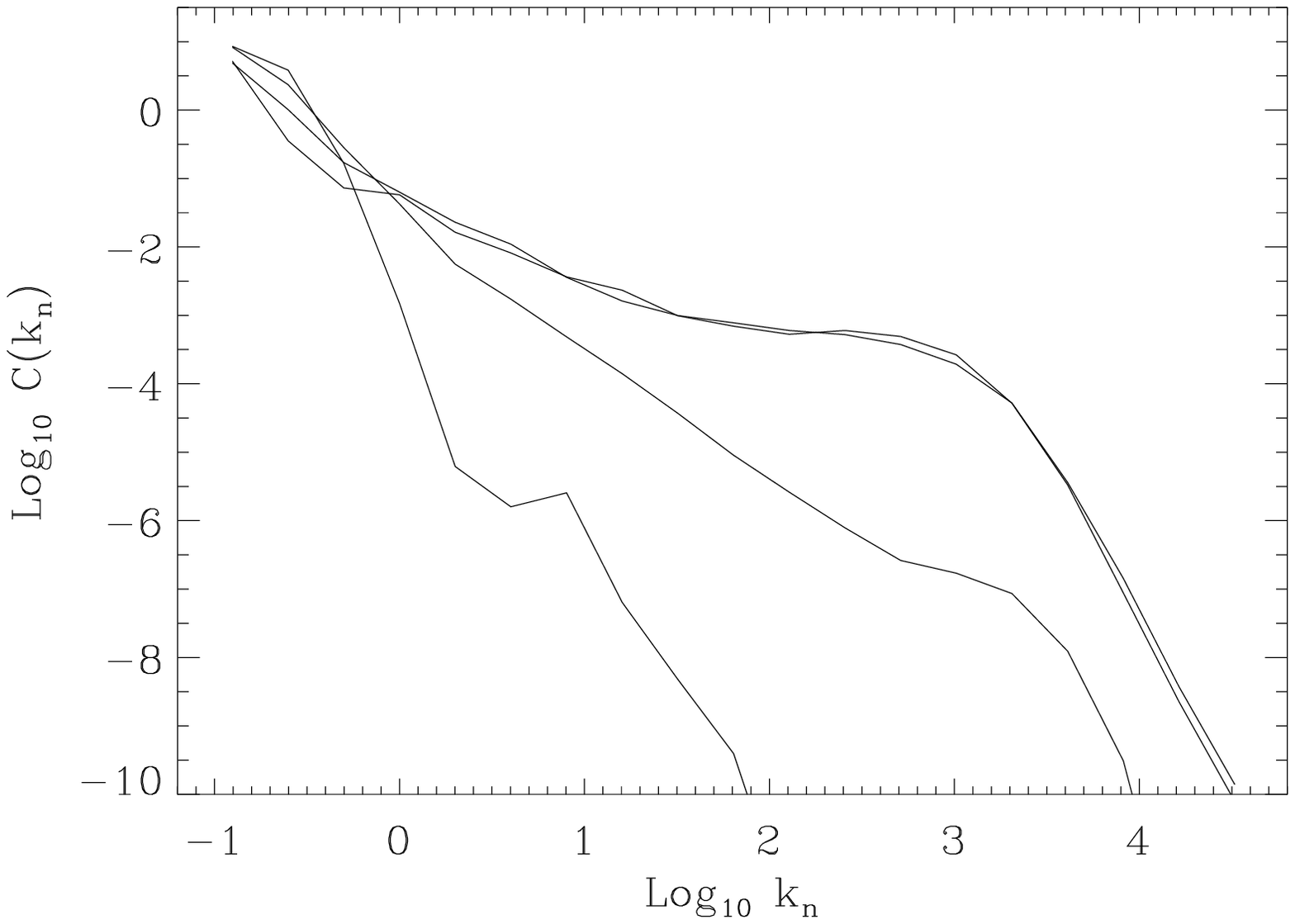}}
\vspace*{.5cm}
\hspace*{.1cm}
{\epsfxsize=3.2in \epsfysize=3.2in \epsfbox{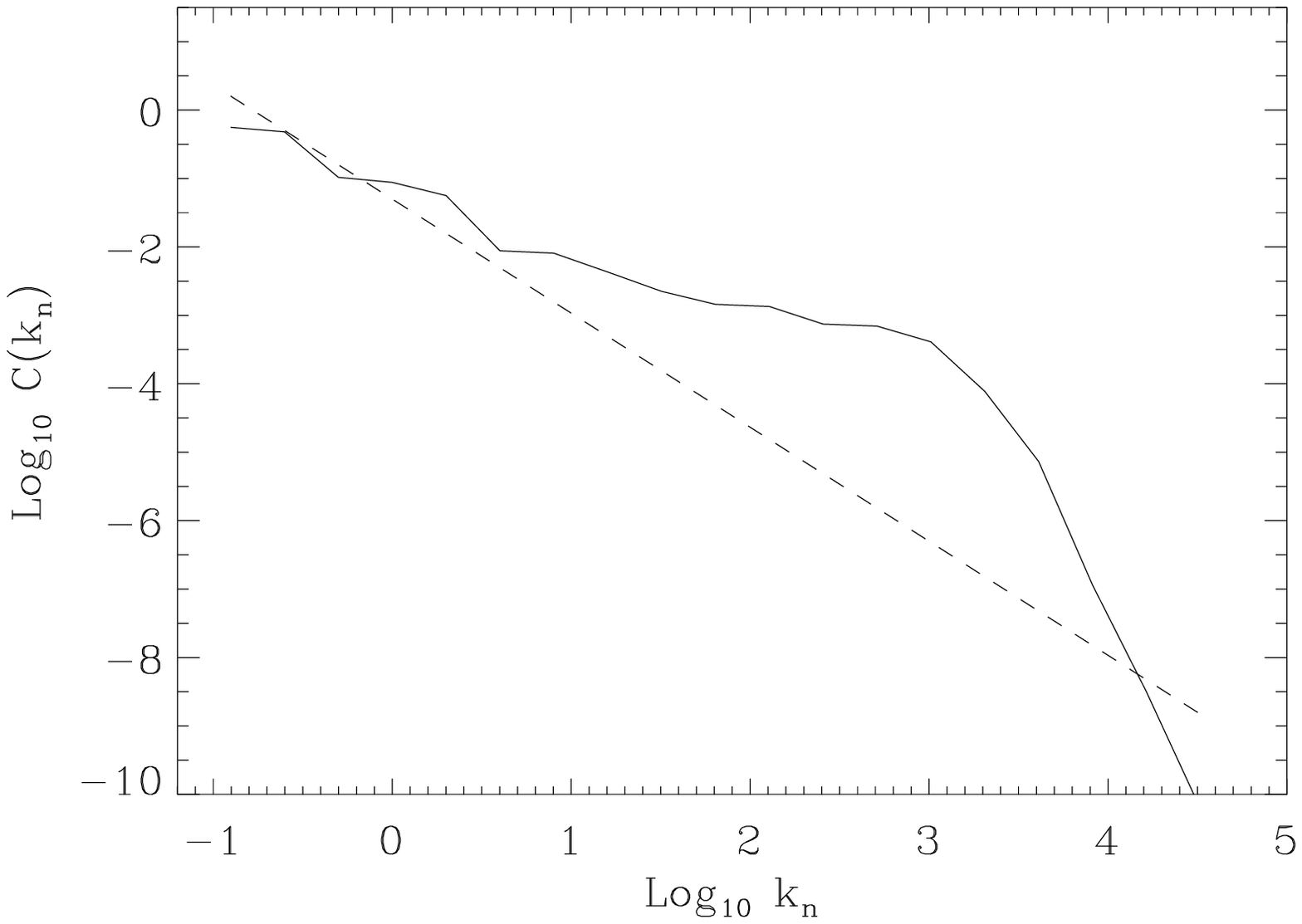}}}
\vspace*{-1cm}
\hspace*{3cm}
\caption[ggg]{
To the left, spectra $C(k_n)$ obtained from numerical integrations
of the shell model, with parameters 
$N$ = 19, $k_0 $= 2$^{-4}$, $\nu$ = 10$^{-6}$, $D$ = 10$^{-13}$,
$\alpha$ = 10$^{-12}$. The different spectra are averaged over 4 time units
and the distance between the spectra are $\sim$ 50 time units. Time
progresses from the left to the right. The right hand figure
shows one particular spectrum. The dashed line has slope -5/3.}
\label{fm2a}
\end{figure}

\begin{figure}[htb]
\centerline{\epsfxsize=3.5in \epsfysize=3.5in \epsfbox{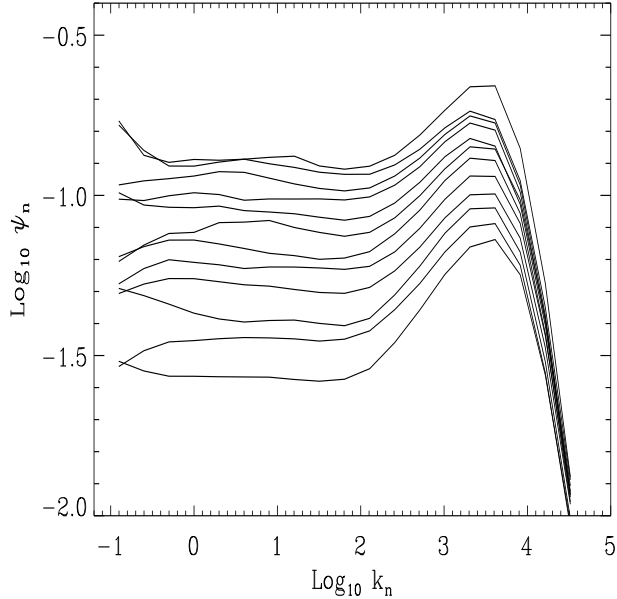}}
\caption[ccc]{A plot of $< \mid \psi_n \mid >$ versus $k_n$ on logarithmic
scales. The parameters are $N$ = 19, $k_0$ = 2$^{-4}$, $\nu$ = 10$^{-6}$, 
$D$ = 10$^{-12}$, and $\alpha$ = 10$^{-12}$. Each curve is averaged over 800
time units and the time intervals between the curves are $\sim$8000 time
units. Time progresses from the uppermost curve to the lower.}
\label{fm3}
\end{figure}

Fig. \ref{fm2a}a shows the development of the spectrum
$C(k_n)$ for a lower value of the viscosity, $\nu = 10^{-6}$ corresponding
to $Re \sim 10^6$. In this case $D = 10^{-13}$ leading
to $Pr = 10^7$. Again, starting from an initial condition 
concentrated on the small $k_n$-values, one observes the occurrence
of the peak at large $k_n$-values. Fig. \ref{fm2a}b shows one of the
spectra and the dashed line corresponds to $C(k_n) \sim k_n^{-5/3}$.
In order to get a more clear picture of the peak, we 
plot  $< \mid \psi_n \mid >$ versus $k_n$ on logarithmic scales. 
For a usual Obukhov spectrum one should find
$< \mid \psi_n \mid > \sim k_n^{-1/3}$, whereas for the viscous-convective 
regime $< \mid \psi_n \mid > \sim$ const. and the peak will therefore appear 
more pronounced.
Fig. \ref{fm3} shows the time development of the
value of the scalar, where each curve is averaged over 800 time units
(the time intervals between the curves are $\sim$8000 time units).
One observes that $\psi_n$ is almost constant over the leftmost regime of the
spectrum, corresponding to $C(k_n) \sim k_n^{-1}$, whereas the peak is
situated at the rightmost part of the spectrum. The peak slowly decreases in
intensity (at a rate determined by the mixing time) but for high values of
the Prandtl number this decrease occurs extremely slowly: the peak in
Fig. \ref{fm3} diminishes significantly only after about $10^6$ time units. 
Simultaneously, the field as a whole slowly vanishes,
$\psi_n \to 0$. This is of course related to the fact that
the passive scalar equation \rf{m10} is not forced and the input
to the motion of $\psi_n$ is only driven by the advective term
which includes $u_n$. Nevertheless, as $Pr\to 0$, $\psi_n \to 0$
slowly, which indeed is reflected in the corresponding value of the mixing
time. The presence of a peak appears to be independent of the initial
state; one can either, as in Figs.\ref{fm1a},\ref{fm2a}, apply an 
initial disturbance which is concentrated at the small $k_n$ values,
or choose states of $\psi_n$ and $u_n$ which are 
solutions to the passive scalar equations, i.e. in which $\alpha =0$.
In all cases the results is a peak in $\psi_n$ at large $k_n$,
so an initial perturbation at the large scales in not a necessity in order
to observe the enhanced ``delay" in the mixing of the two fluids.

In comparison with previous work \cite{Ruiz1} and \cite{Ruiz2}, it should be
emphasized that in our case the peak is {\sl orders of magnitude smaller} in
height. In Ref. \cite{Ruiz2}, the peak (after a short time) has a height
of order some decades. In our case, as seen from the various figures,
the height is only of order half a decade at most, and in many cases it
is much less, which could make experimental observations difficult,
as indeed appears to be the case \cite{Walter2,Walter4}.

\begin{figure}[htb]
\centerline{\epsfxsize=3.5in \epsfysize=3.5in \epsfbox{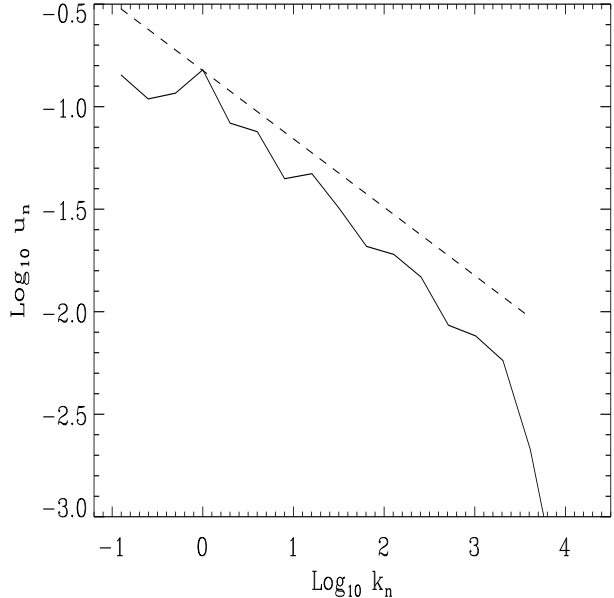}}
\caption[uuu]{The velocity spectrum $\mid u_n \mid$
versus $k_n$ for the same parameters as in
Fig. \ref{fm3}. The spectrum is averaged over $\sim$50000 time units. 
The dashed line has a slope -1/3.}
\label{fm4}
\end{figure}

The corresponding velocity spectrum does not show any sign
of a peak, Fig. \ref{fm4}. On the contrary,
the cut-off at high wavenumbers in the spectrum in fact seems to move
to shorter wavenumbers (larger scales) as a results of the active coupling.
The oscillations in the spectrum is on the other hand much more pronounced
than for the usual ``GOY'' model. This is related to the fact 
that the motion for the present model is also strongly
intermittent and leads to corrections to the Kolmogorov theory. A study of
the intermittency effects is reserved to a forthcoming publication.

\section{The theoretical predictions by Ruiz and Nelson.}

Ruiz and Nelson \cite{Ruiz1,Ruiz2} have proposed a theory for the
dependence of the mixing time on the hydrodynamical parameters.
In the case of the passive scalar, $\alpha = 0$, there is not
a peak in the spectrum but an inhomogeneity created at the
large scale will still persist for a mixing time which is composed of
three terms
\beq
\tau_{pass} = t_0  + {\rm ln} (k'_d  / k_d) + {1 \over 
{D(k'_d  )^2}}~~.
\label{m19}
\eeq
Here, the first term is the time it takes for a perturbation created at
the large scale to reach the dissipative wavenumber $k_d$
(the Kolmogorov length). The second term is the time it takes for a
perturbation to go from the dissipative wavenumber, through the
``viscous-convective'' regime down to the Batchelor wavenumber $k'_d
= k_d Pr^{1/2}$, and the last term is the time actually needed to dissipate
the disturbance at $k'_d$. The second term is found to be
of the order ${{{\rm ln} Pr} \over {2 Re^{1/2}}}$.
For large values of $Re$, the mixing time is therefore of the order $t_0$,
unless ${\rm ln} Pr \gg Re^{1/2}$ is which case the second term dominates.
In the case of an active scalar on the other hand, $\alpha \not=0$, a peak
occurs in the spectrum at
a specific wavenumber $k^*$, and the last term will therefore 
dominate when $Pr \gg Re$. The corresponding mixing time is \cite{Ruiz2}
\beq
\tau_{active} \simeq {1 \over {D(k^*)^2}} \simeq t_0 {Pr \over Re}
\label{m20}
\eeq
Also, the wavenumber of the peak is predicted to be located 
at $k^* \simeq k_0 Re$,
where $k_0$ is the wavenumber of an initial perturbation. We observe
from this theory, that for large values of the Prandtl number,
the mixing time is much longer in the active case than in the passive case.

\begin{figure}[htb]
\centerline{\epsfxsize=3.5in \epsfysize=3.5in \epsfbox{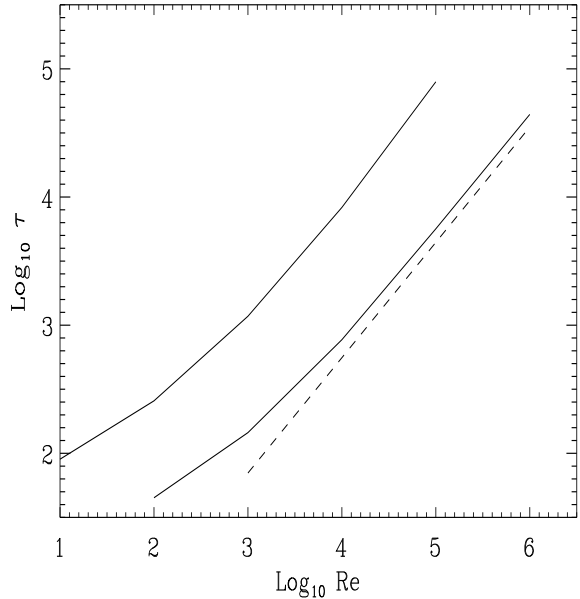}}
\caption[ppp]{The mixing time $\tau_{active}$ versus the Prandtl number $Pr$
estimated from numerical calculations of shell model with the 
following parameters: Upper curve: $N$ = 16, $k_0$ = 2$^{-4}$,
$\nu$ = 10$^{-4}$,~$\alpha$ = 10$^{-8}$ and the gate
$\psi_G$ = 10$^{-6}$. Lower curve: $N$ = 19, $k_0$ = 2$^{-4}$,
$\nu$ =10$^{-6}$, $\alpha$ = 10$^{-12}$, and 
the gate $\psi_G$ = 10$^{-4}$. The slope of the dashed line is 0.9.}
\label{fm5}
\end{figure}

We have tested the prediction of the theory given by Eq.\rf{m20}
using the shell model introduced in the previous sections. Fig. \ref{fm5}
shows a plot of the mixing time $\tau_{active}$ versus the Prandtl
number for two different values of the Reynolds number. In the first case,
$Re \sim 10^6$ and $Pr$ in the range from $10^2$ to $10^6$ and in the
second, $Re \sim 10^4$ and $Pr$ in the range $10$ to $10^5$. The
mixing time is estimated in the following way. At the shell corresponding
to the $k_n$-value on which the maximum of the peak is localized, we monitor
the value of $< \mid \psi_n \mid >$, where the average is over a specific time
interval. As this value decreases below a chosen gate $\psi_G$, the
corresponding time is associated with the mixing time $\tau_{active}$, 
at that particular value of the Prandtl number.
Then the value of $Pr$ is changed and starting from the same initial
conditions, using the same value of the gate, one obtains the new value
of $\tau_{active}$, and so on. The 
mixing time clearly diverges with the Prandtl number as a power law,
$\tau_{active} \sim Pr^{\beta}$. The best fit to the data produces a value
$\beta \sim 0.9$ for $Re \sim 10^6$ (as indicated by the dashed line)
and $\beta \sim 0.95$ for $Re \sim 10^4$. The prediction \rf{m20}
of Ruiz and Nelson is valid in the limit $Pr \gg Re$ and this only holds for
our data where $Re \sim 10^4$, so the agreement with the theory is reasonable.
It is however tempting to conjecture that the presence of intermittency
might cause the exponent $\beta$ to decrease below 1. The shell model
presented here in Eqs. \rf{m10} and \rf{m12} exhibits strongly intermittent
motion where the laminar periods are interrupted by bursts of
violent motion. The higher the Reynolds number, the more pronounced is the 
intermittency and since the largest deviations from the 
prediction \rf{m20} is observed at $Re \sim 10^6$, the effects of
intermittency could diminish the value of mixing time due to the presence of
the long laminar periods.

Goldburg and co-workers investigated in several experiments the possible 
existence of the active coupling term in binary mixtures. In Ref.
\cite{Walter1} the presence of an active term was expected
because the mixing time was found experimentally to increase 
dramatically with the value of the Reynolds number. Nevertheless,
the experimental data did not follow the prediction \rf{m20}.
Later experiments showed, however, that the long mixing times were
caused by the fact that stirring of the binary mixture cools it down below
the critical temperature, into the region where the system is
immiscible and phase separation is favored \cite{Walter2}. Subsequent
measurements on phase separations and the correlation functions
of temporal fluctuations did not show any sign of an anomalous peak
in the spectra \cite{Walter3,Walter4}. In Ref. \cite{Walter4} it was
already argued that intermittency effects might strongly influence
the critical fluctuations leading to correlation functions which
are stretched exponentials. That intermittency effects are very important 
close to the critical point is in accordance with the results presented 
in this paper. Conclusively, one must therefore say that, in spite of
several experimental attempts, there is no clear evidence for the active
coupling term in \rf{m6}. We return to this point in sections 7 and 8.

\section{The continuous cascade model for hydrodynamics}

The Kolmogorov scaling behavior is a static solution of the energy cascade
model. In this section we discuss time-dependent generalizations of the
simple $k^{-5/3}$ Kolmogorov behavior of the power spectrum, using a
continuous version of the cascade model first discussed by
Parisi \cite{parisi}. Initially these spectra start out as an arbitrary
power behaviour $k^{2p-1}$, where $p$ is some constant which
can be selected as one wishes (if one includes necessary cutoffs in
$k$-space), but it turns out that after a short time the
large $k$ behavior becomes of the Kolmogorov type, with some time
dependence. Thus, at a given time the spectra have a ``two-slope''
structure with a smooth interpolation. The solution is only valid in the
inertial range, where diffusion can be ignored.

The simplest hydrodynamical cascade model is given by the equation of
motion \cite{Ok-Ya},
\beq
(du_n/dt+\nu k^2u_n)^*=-ik_n(u_{n+1}u_{n+2}-\frac{\delta}{r}u_{n-1}u_{n+1}
-\frac{1-\delta}{r^2}u_{n-1}u_{n-2})+F_n,
\label{1} 
\eeq
where $F_n$ represents an external forcing. From the point of view of energy
(i.e. $\sum |u_n|^2$) conservation the parameter $\delta$ is arbitrary,
but it can be fixed by requiring conservation of (generalized) helicity
\cite{kadf}. In our case we keep the parameter $\delta$ arbitrary. 

Some years ago Parisi \cite{parisi} studied eq. \rf{1} in the limit
\footnote{In ref. \cite{Ruiz1} this limit was considered for a different
type of shell model. The resulting equation is linear, in contrast to those
considered in this paper.}
$r\rightarrow 1$, meaning that the distance between the shells goes to zero. 
Taking $r=1+\epsilon$ it is easily seen that one gets
\beq
\left({\pa \over\pa t}+\nu k^2\right)u^*=-i\epsilon (2-\delta)k\left(u^2+3ku
{\pa u\over\pa k}\right)+F(k)+ O(\epsilon^2).
\label{2}
\eeq
One might consider this equation to be a ``model of a model'', and thus very
academic. However, as we shall see, this model and, in particular, its
generalizations, satisfy the same conservation laws as the corresponding
discrete models. Therefore one can equally well consider the continuous
version as a model in its own right, with the advantage that it is
considerably simpler than the discrete versions.

We can now scale $\epsilon (2-\delta)$ into time $t$ and then let
\footnote{There exists the
possibility of taking the more exotic limit $|\epsilon (2-\delta)|\rightarrow$
a finite value, so that $|\delta|$ approaches infinity.}
$\epsilon\rightarrow $ 0. As just mentioned, the resulting model can be
considered as being independent of the discrete version, since it satisfies
the relevant conservation laws. Instead, one can consider eq. \rf{2} to be an
approximation to the discrete model, to be supplemented by higher order terms
in $\epsilon$ if needed. 

We shall study eq.\rf{2} in the inertial range, where viscosity can be
ignored\footnote{For the special case $p$=1 we shall include diffusion later
in this section.}. Also, we disregard the possible forcing term, so it is then
clear that the motion must die out after some time if diffusion is
included. Hence we study
the equation
\beq
{\pa u^*\over\pa t}=-ik\left(u^2+3ku{\pa u\over\pa k}\right)+O(\epsilon).
\label{3}
\eeq
We now choose a special phase and make the ansatz $u=ik^pf(k^qt)$.
Inserting this in eq. \rf{3} we obtain $q=1+p$, i.e.
\beq
u=ik^pf(k^{1+p}t).
\label{4}
\eeq
This scaling was first considered by Parisi \cite{parisi}, except for the
special case $p$=1, which was introduced many years ago by Heisenberg
\cite{heisenberg}. It solves the discrete as well as the continuous model.

A scaling of the type \rf{4} is consistent with the
well known invariance of the Navier-Stokes equations (see for instance 
\cite{frisch}),
\beq
l\rightarrow \lambda l,~u\rightarrow \lambda^h u,~t\rightarrow \lambda^{1-h}t,~
{\rm and}~\nu\rightarrow\lambda^{1+h}\nu,
\label{self}
\eeq
in the inertial range, where we can take $\nu$=0. The reason for this is
that the self-similarity \rf{self} can be translated to $k$-space
with $l\rightarrow 1/k$, and it then corresponds to the scaling \rf{4}
with $p=-h$. The main point is that the scaling variable $k^{1+p}t$ is then an
invariant. Also, in the scaling \rf{self} $l$ is usually interpreted as 
the scale of an eddy, and the typical velocity of this eddy is then
$|\vek u(\vek x+\vek l)-\vek u(\vek x)|$. This compares
excellently to the velocity mode $u_n$ used in $k$-space, where $u_n$ is
the velocity increment over an eddy of scale $l\sim 1/k_n$. 
Let us further note  that for $p=-h$=1, diffusion can be included, since
\rf{self} then leaves $\nu$ invariant.

It should be emphasized that the expression \rf{4} has the explicit power $k^p$
in front to accommodate the self-similarity transformation \rf{self} for
$u$. However, the function $f$ depends only on the quantity $k^{1+p}t$,
which is {\sl invariant} under the self-similarity transformations
\rf{self}. In the absence of a solution for $f$, this function can be
completely arbitrary from the point of view of self-similarity, and hence
e.g. the time evolution cannot be predicted at all. This implies that {\sl
a priori} there is {\sl no} agreement with K41 theory (see \cite{frisch}
for a general discussion of K41). Of course, having a 
solution for $f$ changes the situation.   
 
Eq. \rf{4} thus means that the velocity mode is initially assumed to be of
the form $k^p$. Physically, one can imagine that this initial condition is
produced by some external force. In this sense the selection of initial
conditions is equivalent to initial forcing. Also, $p$ governs the
initial correlation function $<u_i(\vek x)u_k(\vek y)>$. The cases
$p$=3/2 or 1 correspond to Gaussian disorder in three and two dimensions,
respectively, i.e. $<u_i(\vek x)u_k(\vek y)>\propto
\delta_{ik}\delta^3(\vek x-\vek y)$ or $\delta_{ik}\delta^2(\vek x-\vek y)$,
respectively\footnote{Here we leave out a discussion of the consequences of
div$\vek u$=0, which requires a projection operator in the definition of
Gaussian randomness. This is of no relevance in the following.}. This is 
because the k-space energy spectrum is given by
\beq
E(k,t)=|u(k,t)|^2/k={\rm const}~k^{D-1}\int d^D x \exp (i\vek k\vek x) <\vek u
(\vek x,t)\vek u(0,t)>,
\label{fourier}
\eeq
in $D$ dimensions.

In order to have a convergent energy
\beq
E=\int dk |u|^2/k,
\label{4a}
\eeq
we obviously need a ultraviolet cutoff for $p\geq$1/2. Similarly, for $p\leq$0 
an infrared cutoff is needed.

At this stage we need a discussion of the boundary conditions associated with
\rf{2} and \rf{3}. From \rf{2} we get energy conservation in the absence of
forcing and viscosity provided 
\beq
k|u(k,t)|^3\rightarrow 0~ ~{\rm for}~~ k\rightarrow 0~~ {\rm and}~~\infty.
\label{kuk}
\eeq
If this condition\footnote{In the discrete version
\rf{1} there is, strictly speaking, a similar boundary condition if $n$ goes
to infinity. This is because when one checks energy conservation, sums of the
type $\sum^\infty  k_n uuu$ (the $u$'s have different indices) are
encountered. Although there is a complete cancellation of these terms, the sums
only exist in a strict mathematical sense if all terms of the
type $k_n uuu$ vanish for $n\rightarrow\infty$.} is not satisfied for
$k\rightarrow \infty$  there is ``diffusion at infinity''\cite{parisi}. 

Inserting \rf{4} in eq.\rf{3}, we obtain the following equation for $f$,
\beq
\frac{df(x)}{dx}=-\frac{(1+3p)f(x)^2}{1+3(1+p)xf(x)},~~x=k^{1+p}t.
\label{5}
\eeq 
This equation can be simplified by the substitution
\beq
f(x)=g(x)/x,
\label{6}
\eeq
and we get
\beq
\frac{dg(x)}{d \ln x}=\frac{g(x)+2g(x)^2}{1+3(1+p)g(x)}.
\label{sveske}
\eeq
The solution is given by
\beq
g(x)(1+2g(x))^{(1+3p)/2}=x/x_0.
\label{7}
\eeq
Here $x_0$ is an arbitrary constant which gives the strength of the initial
velocity mode,
\beq
u(k,0)=ik^p/x_0.
\label{8}
\eeq
When $x\rightarrow\infty$ we find from \rf{7} that
\beq
g(x)\rightarrow 2^{-(1+3p)/3(1+p)}~(x/x_0)^{2/3(1+p)}.
\label{9}
\eeq
Inserting this in \rf{4} and \rf{7} we get
\beq
|u(k,t)|\rightarrow 2^{-\frac{1+3p}{3(1+p)}}~x_0^{-\frac{2}{3(1+p)}}~
t^{-\frac{(1+3p)}{3(1+p)}}~k^{-\frac{1}{3}}.
\label{10}
\eeq
Thus we see that irrespective of the initial spectrum \rf{8} the velocity
approaches the Kolmogorov spectrum with a time dependent amplitude for large
values of $k$ and/or time. Note that this decay law does not agree with 
the classical theory put forward by Karman, Howarth and Kolmogorov
which is of course due to the non-triviality of the function $f$ in
\rf{4} \cite{karman,Kol} (see the general discussion on classical
results on decay laws by Frisch \cite{frisch}). 

There are a few special cases where eq.\rf{7} can be solved explicitly. The
simplest is the case where $p=-1/3$, where we get $g(x)=x/x_0$, leading
to the time independent Kolmogorov spectrum,
\beq
|u(k,t)|=k^{-1/3}/x_0,~~{\rm for}~~p=-1/3.
\label{11}
\eeq
This result is trivial, since it is easy to see that the original
equation \rf{3} has \rf{11} as a static solution.

A non-trivial result can be obtained by considering the case $p=1/3$, where
eq.\rf{7} becomes second order in $g$. Using \rf{4} and \rf{6}
we then obtain
\beq
|u(k,t)|=\frac{1}{4kt}\left(-1+\sqrt{1+8k^{4/3}t/x_0}\right).
\label{12}
\eeq
For small but non-vanishing $t$, the slope in the
corresponding power spectrum changes from -1/3 to -5/3. For
$k\rightarrow\infty$ the boundary condition \rf{kuk} is not satisfied. This is
simply a consequence of the fact that the Kolmogorov spectrum does not
satisfy this condition. As mentioned before, we only expect the solution
\rf{12} to be relevant for the inertial range, where diffusion is negligible.
For completeness we mention that the energy integrated to $k=\infty$ is given
by
\beq
E=\sqrt{2}/(x_0^2\sqrt{t}).
\label{14}
\eeq
The energy contents approaches zero as time goes on simply because the class of
solutions \rf{4} correspond to an inverse cascade moving towards smaller
values of $k$, ultimately reaching the value 0 except in the point $k=0$,
where it is given by 1/$x_0^2$. In ``$\vek x$-space'' this means that small
scale structures in time become structures of infinite extension, with no
local energy. This shows up if we consider the ``integral scale'' $l_0$ used
in turbulence theory as a measure of possible large scale structures,
\beq
l_0=\int\frac{dk}{k} E(k)\left(\int dk E(k)\right)^{-1},~{\rm with}~
E(k)=|u|^2/k.
\label{is}
\eeq
This quantity behaves like $t^{3/4}$, and in the general case it goes as
$t^{1/(1+p)}$. Hence, for large times the structures become very extended.

Another case which can be solved is $p=1$, since eq.\rf{7} then becomes a
cubic equation for $g$. We find
\beq
|u(k,t)|=R(s)/kt,
\label{sidstenyt}
\eeq
where
\beq
R(s)=\frac{1}{6}\left(\left(1+s \left(1+
\sqrt{1+\frac{2}{s}}\right)\right)^{\frac{1}{3}}+\left(1+
s \left(1+\sqrt{1+\frac{2}{s}}\right)\right)^{-\frac{1}{3}}-2\right),
\label{15}
\eeq
and
\beq
s=27k^2t/x_0.
\label{extra}
\eeq
The energy spectrum initially has the slope +1. For small $k$ the slope
remains +1, but at larger values of $k$ the slope turns into -5/3, in
accordance with the universality of the Kolmogorov spectrum at large
$k$'s. Again there is an inverse cascade moving the energy towards
smaller $k$'s as time passes.  

\begin{figure}[htb]
\vspace*{-3cm}
\mbox{{\epsfxsize=3.2in \epsfysize=5in \epsfbox{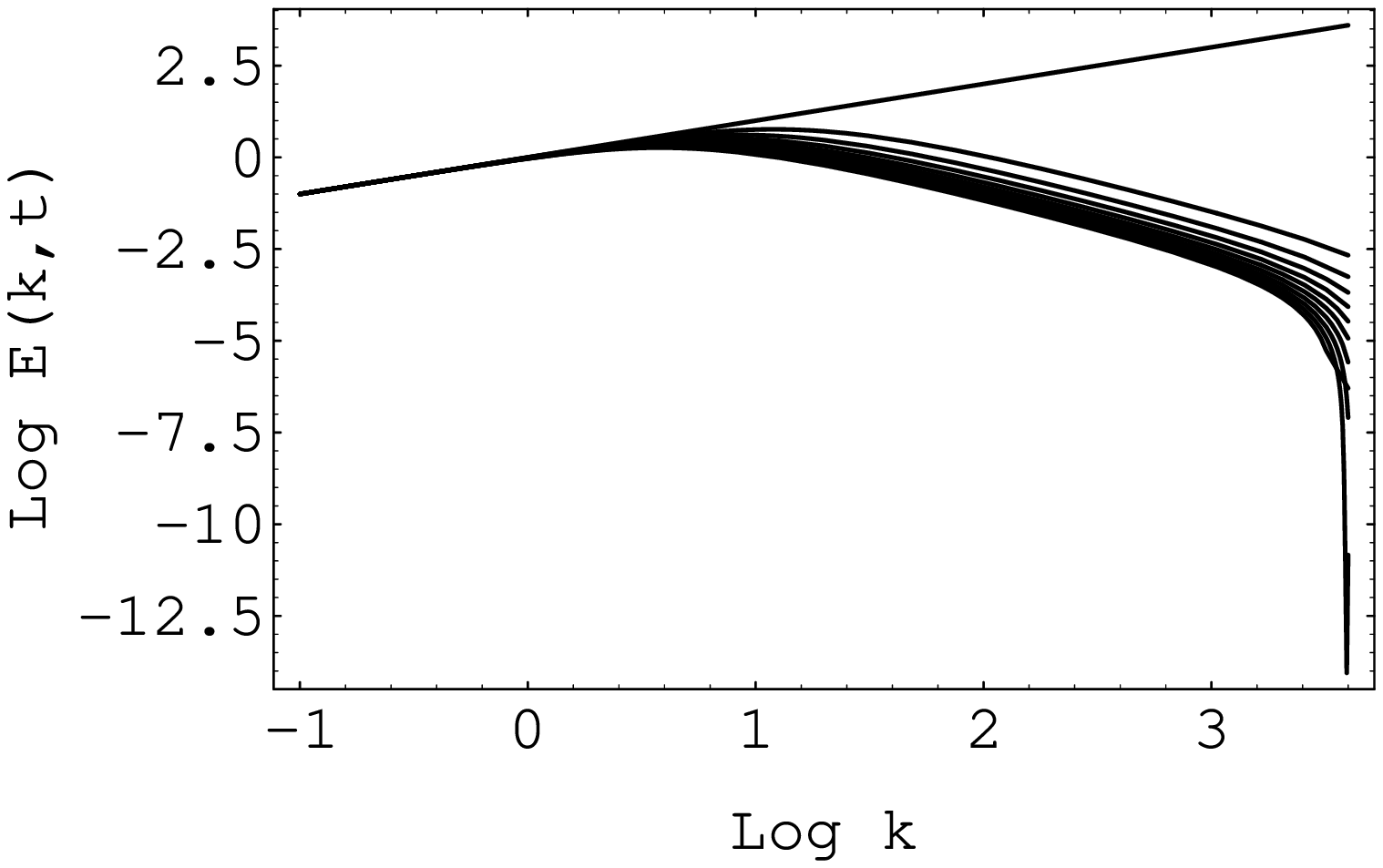}}
\vspace*{.5cm}
\hspace{.1cm}
{\epsfxsize=3.2in \epsfysize=5in \epsfbox{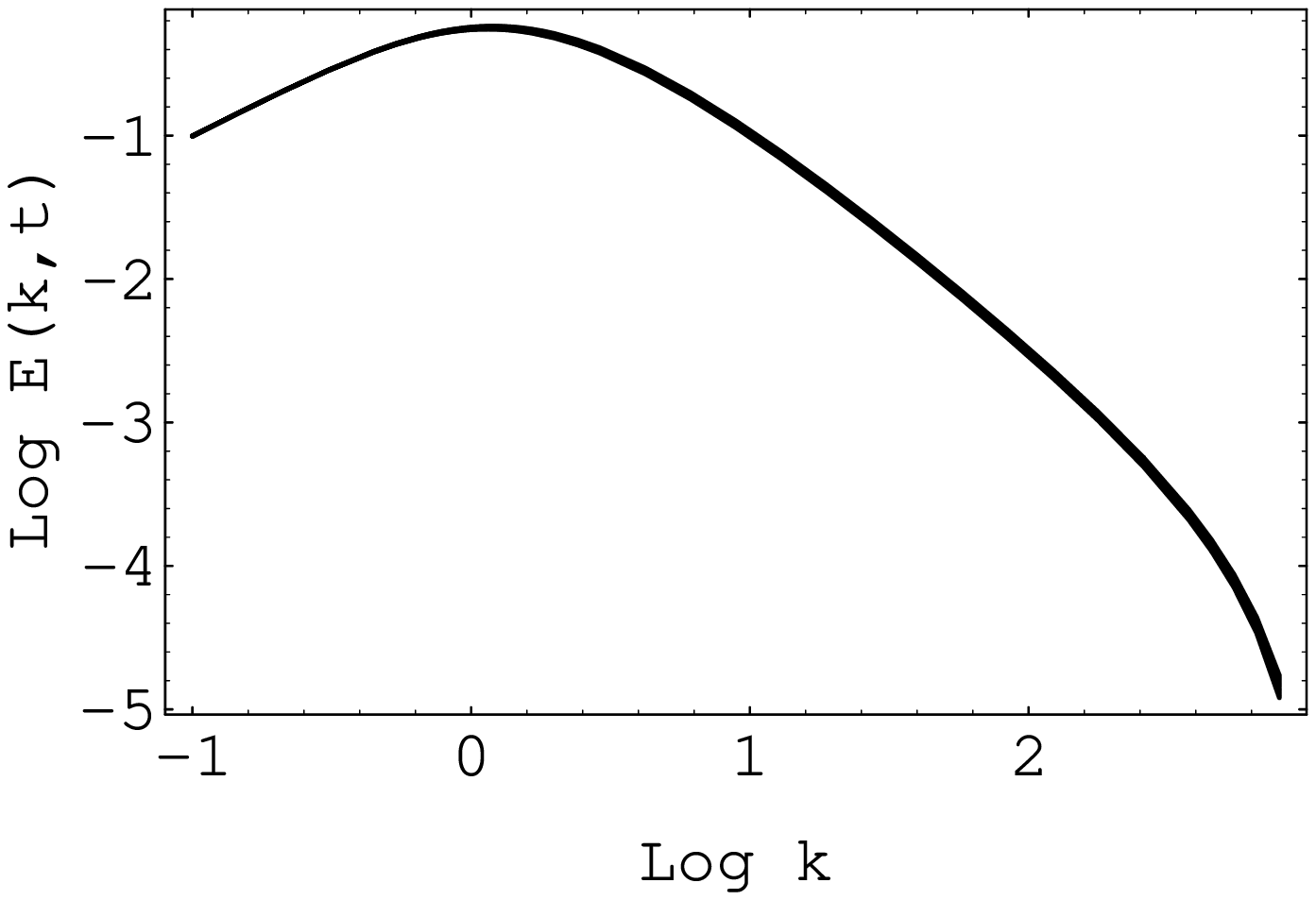}}}
\vspace*{-4cm}
\caption[yyy]{A double logarithmic (base 10) plot of the energy spectrum
$E(k,t)$ as a function of $k$ in the continuous model. In the left hand figure
we show ten small values of time, from $t$=0 (where the $E$ goes like $k$) to
$t$=10$^{-2}$. In the right hand figure we show $E$ for ten values of time 
between $t$=1 and $t$=1.1. In both cases time moves from the right to
the left. The viscosity is $\nu$=10$^{-2}$, $x_0$
=1, and from \rf{cut} log$_{10} k_D$ is of order 2.5 (left) and 2 (right),
in reasonable agreement with the figures.}
\vspace*{0cm}
\hspace*{3cm}
\label{XXX}
\end{figure}

The $p$=1 case is interesting from the point of view of studying the effect
of diffusion. The main feature is that in general the ansatz \rf{4} is not
consistent with the viscosity term on the left hand side of eq. \rf{2}.
However, when $p$=1, the powers of $k$ nicely divide out on both sides of
eq.\rf{2}, leaving them as functions of the scaling variable $x$ only.
Instead of eq.\rf{5} we obtain
\beq
\frac{df(x)}{dx}=-\frac{4f(x)^2+\nu f(x)}{1+6xf(x)}.
\label{dif}
\eeq
The substitution \rf{6} cannot be used to solve this equation because of
the dissipative term. However, eq.\rf{dif} can of course easily be solved
numerically, and the result compared to the analytic solution \rf{15}.
To see the effects expected, let us compute the diffusion cutoff defined by
$f^2\sim\nu f$ (see eq.\rf{dif}), using eqs. \rf{4},\rf{6}, and \rf{9} applied
to the case $p$=1, 
\beq
x_D\sim x_0^{-1/2}~\nu^{-3/2},~~ {\rm or}~~k_D\sim \nu^{-3/4}~x_0^{-1/4}~
t^{-1/2},
\label{cut}
\eeq
which is what one one expects as far as the dependence on $\nu$ is concerned.
However, the time dependence should be noticed. The
latter reflects the fact that we have an inverse cascade. The cutoff \rf{cut}
is in qualitative agreement with what one obtains by comparing the analytic
solution \rf{15} with the numerical solution of \rf{dif}. In fig.~\ref{XXX}
we give an example of a solution, where we plot the energy. It is clearly seen
that there is a change from the initial slope +1 to -5/3. 

For a general $p$ the result \rf{cut} can be extended to
\beq
k_D\sim \nu^{-3/4}~x_0^{-1/2(1+p)}~t^{-(1+3p)/4(1+p)}.
\label{p}
\eeq
The behavior of $k_D$ as a function of $\nu$ is thus universal, i.e.
independent of $p$, whereas the time dependence of $k_D$ is clearly
non-universal. 

The results \rf{cut} and \rf{p} are to some extent consistent with Kolmogorov's
scaling arguments as far as the $\nu$-dependence $and$ as far as the
time dependence are concerned: The energy should go as $E(k)\sim
\epsilon^{2/3}k^{-5/3}$, where $\epsilon$ is the kinetic energy per
unit mass (anyhow put equal to one) and time, with $\epsilon=dE/dt\sim ku^3$
for large $k$. Now from \rf{10} $ku^3\sim t^{-(1+3p)/(1+p)}$ for large $k$.
{}From Kolmogorov's arguments one expects the dissipation scale $k_D\sim 
\nu^{-3/4}~\epsilon^{1/4}$, i.e.
$k_D\sim \nu^{-3/4}~t^{-(1+3p)/4(1+p)}$. This is exactly the result
\rf{p}, also as far as the time dependence is concerned. Of course, the fact 
that $\epsilon$ is not a constant in our case, does not conform to Kolmogorov's
argument.

In the case $p=$1 it is possible to obtain an implicit equation for $g(x)$
even when diffusion is included. For $\nu\neq$0,
proceeding like in eqs.\rf{5},\rf{6},\rf{sveske},\rf{7}, and \rf{15}, we get
that eq. \rf{7} is replaced by,
\beq
g(k^2t)(1+2g(k^2t))^2=\frac{k^2t}{x_0}\exp\left(-\nu\int_0^{k^2t}dx
\frac{1}{1+2g(x)}\right).
\label{diffy}
\eeq
Therefore eq. \rf{sidstenyt} is changed to
\beq
|u(k,t)|=R(d)/kt,
\label{allersidstenyt}
\eeq
where $R(d)$ is given by \rf{15}, and where
\beq
d=\frac{27k^2t}{x_0} \exp\left(-\nu\int_0^{k^2t}dx\frac{1}{1+2g(x)}\right).
\label{heltny}
\eeq
This equation can be used iteratively, starting from the behaviour for
$\nu$=0. To the lowest non-trivial order in $\nu$ eq. \rf{allersidstenyt}
becomes
\beq
|u(k,t)|=R\left(27k^2t\exp(-\nu k^2t)/x_0\right)/kt.
\label{diffa}
\eeq
for $k^2t$ small. It is interesting that \rf{diffa} is a self-consistent
solution of eq. \rf{diffy} for $k^2t$ small and large. Thus the decay
of the velocity at large $k$ is exponential.  For $k^2t$ neither small
or large, the expression given by \rf{15} and \rf{heltny}
indicates that an exact inclusion of diffusion is rather complex, even in
this simple model.

It is very interesting that the scaling behaviour for the case $p$=1, i.e.
$u(k,t)=kf(k^2t)$ was first considered by Heisenberg in his model, where in
the range from 0 to $k$, the action of all smaller eddies are assumed to
be represented by an effective viscosity \cite{heisenberg}. Although it is not
obvious that the cascade model satisfies this assumption, the Heisenberg
scaling appears as a solution. Physically the $k^2t$-scaling can be understood
\cite{heisenberg} by assuming that the spectrum is determined by one length
only, namely the length $\sim 1/k_0$ of the largest eddies. Let the
velocity of these eddies be $v_0$.  On dimensional grounds it then follows
that \cite{heisenberg}
\beq
\frac{d}{dt}(1/k_0)\sim v_0,~{\rm and}~\frac{d}{dt}(1/v_0)\sim k_0,
\label{heisenb}
\eeq
from which one gets $k_0\sim v_0\sim 1/\sqrt{t}$. This is exactly the scaling
in the $p$=1 case. However, the continuous model has an exponential decay
for large $k$, in contrast to Heisenberg's model, which has a power 
behaviour, presumed to be unrealistic \cite{mogens}. 

The solutions discussed above refer to the case when the phase is fixed. One
can try to find solutions where the phase plays a dynamical role by making
the scaling ansatz $u=ik^pf(k^{1+p})$ with $f$ a complex function, in which
case eq. \rf{5} is replaced by
\beq
\frac{df^*(x)}{dx}=-3(1+p)xf(x)\frac{df(x)}{dx}-(1+3p)f(x)^2.
\label{complex}
\eeq 
For the case $p$=1/3, introducing again the substitution \rf{6} with $g(x)$
complex, this leads to the result
\beq
g^*(x)+2g(x)^2=x/x_0,~~x=k^{4/3}t.
\label{complex3}
\eeq
This equation has the previously discussed real solution corresponding to
\rf{12}, as well as a new complex solution with
\beq
{\rm Re}~g(x)=1/4,~~{\rm Im}~g(x)=\sqrt{(3-8x/x_0)}/4.
\label{complex2}
\eeq
This solution is obviously only valid for $x \leq 3/8~x_0$. Since $x$ can take
any value from 0 to $\infty$ the complex solution should be rejected.

To summarize the results obtained so far, one can say that the
continuous model gives interesting, non-trivial results. This model is
most interesting for the case where $p$=1, since then diffusion can
be included. The initial condition $|u(k,0)|\propto k$ then corresponds
to an initial Gaussian disorder in two dimensions. Therefore we believe
that this model may be of most relevance in two dimensions. This is
consistent with the fact that the model has an $inverse$ cascade.

There exists a generalization to a model with helicity, namely the continuous
version of a model introduced by Biferale and Kerr \cite{bif}, leading to
\cite{mogens}
\beq
\left({\pa\over\pa t}+\nu k^2\right)(u^+)^*=-ik\left(4ku^-{\pa u^+\over\pa k}
+2ku^+{\pa u^-\over\pa k}+(2+\alpha)u^+u^--\alpha (u^-)^2\right).
\label{mogens}
\eeq
There is a similar equation with +$\leftrightarrow$-. The energy and the
generalized helicity are conserved,
\beq
E=\int\frac{dk}{k}(|u^+|^2+|u^-|^2),~~H=\int \frac{dk}{k}k^\alpha
(|u^+|^2-|u^-|^2).
\label{helicity}
\eeq
Making the scaling ansatz
\beq
u^+=ik^p~f(k^{1+p}t) ~~{\rm and}~~ u^-=ik^p~h(k^{1+p}t),
\label{ansatz}
\eeq
one obtains from \rf{mogens} by ignoring viscosity
\beq
\frac{df(x)}{dx}=-\frac{2(1+p)xfdh/dx+(6p+2+\alpha)fh-\alpha h^2}
{1+4(1+p)xh}.
\label{skal}
\eeq
There is a similar equation with $f$ and $g$ interchanged. These equations
have a potentially much richer structure than eq.\rf{5}. We also mention
that the discrete and continuous GOY equations have been generalized to
magnetohydrodynamics, using a helicity decomposition \cite{b2}. 

\section{The continuous shell model for turbulent mixtures}

In section 2 we discussed the shell model for binary
mixtures. Proceeding exactly as in the previous section, we can now derive the
continuous version of the relevant equations. Here we shall just give the
results. For convenience we define $\phi=k\psi$, and after a rescaling of
time the equations become
\beq
{\pa \phi^*\over\pa t}+Dk^2\phi^*=ik\left(-v\phi+\phi k{\pa v\over\pa k}+
2 vk{\pa\phi\over\pa k}\right),
\label{aa}
\eeq  
and
\beq
{\pa v^*\over\pa t}+\nu k^2 v^*=ik\left(2\alpha \phi^2+c\left(v^2
+3vk{\pa v\over\pa k}\right)\right).
\label{bb}
\eeq
Here $v$ is the velocity mode and $c$ is an arbitrary constant. These
equations conserve 
\beq
\sum |\psi_n|^2\rightarrow \int \frac{dk}{k}~|\psi|^2,~{\rm and}~~
\sum(|v_n|^2+\alpha |\phi_n|^2)\rightarrow \int \frac{dk}{k}~(|v|^2
+\alpha |\phi|^2),
\label{cc}
\eeq 
provided we have the boundary conditions
\beq
v|\phi|^2/k\rightarrow 0,~k|v|^3\rightarrow 0,~k|\phi|^2v\rightarrow 0,
\label{dd}
\eeq
for $k\rightarrow \infty$ and 0. Again there may be ``diffusion at
infinity'', as discussed in the last section. 

We can now look for scaling solutions of eqs.\rf{aa} and \rf{bb} in the
inertial range. It turns out that they should have the form
(analogously to MHD)
\beq
\phi(k,t)=ik^a~\bar{\phi}(k^{1+a}t),~{\rm and}~v(k,t)=-ik^a~\bar{v}(k^{1+a}t).
\label{ccc}
\eeq
These results have been found by requiring that powers of $k$ on the two
sides of eqs.\rf{aa} and \rf{bb} should cancel out. This fixes the powers in 
$v$ and $\phi$ to be identical. It is interesting that these results are
again consistent with the exact self-similarity of the Navier-Stokes eq.
\rf{self}, supplemented by $\psi\rightarrow \lambda^{-a}\psi$, as can
be seen from eq. \rf{m6}. From eqs.\rf{aa} and \rf{bb} we then get
\beq
-\bar{\phi}'(x)^*=(3a-1)\bar{v}(x)\bar{\phi}(x)
+(1+a)x\bar{\phi}(x)\bar{v}'(x)+2(1+a)x\bar{v}(x)\bar{\phi}'(x)
\label{phi}
\eeq
and
\beq
\bar{v}'(x)^*=-2\alpha\bar{\phi}(x)^2-c((1+3a)\bar{v}(x)^2
+3(1+a)x\bar{v}(x)\bar{v}'(x)).
\label{v}
\eeq
Here the scaling variable $x$ is given by $k^{1+p}t$.

In general the scaling ansatz \rf{ccc} is inconsistent with diffusion, and
hence eqs. \rf{phi} and \rf{v} can only be used in the inertial range.
However, as already seen in the last section, the case $a$=1 is an
exception. In this case all powers of $k$ neatly cancel out even in the
presence of diffusion. If we assume that the functions $\bar{\phi}$
and $\bar{v}$ are real, the equations become
\beq
-\bar{\phi}'(x)/\bar{\phi}(x)=(2\bar{v}(x)+2x\bar{v}'(x)+D)/(1+4x\bar{v}(x)),
\label{phidiffusion}
\eeq
and
\beq
\bar{v}'(x)=-(2\alpha\bar{\phi}(x)^2+4c\bar{v}(x)^2
+\nu \bar{v}(x))/(1+6cx\bar{v}(x)),
\label{vdiffusion}
\eeq
where the scaling variable is now $x=k^2t$.

Eqs. \rf{phidiffusion} and \rf{vdiffusion} can be reformulated in a way
which is similar to what was done in the last section. From \rf{phidiffusion}
we get
\beq
\bar{\phi}(k^2t)=\frac{\bar{\phi }(0)}{\sqrt{1+4k^2t\bar{v}(k^2t)}}\exp\left(-D
\int_0^{k^2t} dx/(1+4x\bar{v}(x))\right).
\label{sol}
\eeq 
Similarly eq.\rf{vdiffusion} can be integrated to give the implicit equation
\beq
\bar{v}(k^2t)(1+2ck^2t\bar{v}(k^2t))^2=\bar{v}(0)\exp\left(-\nu
\int_0^{k^2t}dx\frac{1}{1+2cx\bar{v}(x)}\left(1+\frac{2\alpha}{\nu}
\frac{\bar{\phi}(x)^2}{\bar{v}(x)}\right)\right),
\label{ution}
\eeq
provided $\bar{v}$(0) does not vanish.
This result is similar to eq.\rf{diffy}, and a comparison shows that
the last factor on the right above (containing $\bar{\phi}^2/\bar{v}$)
is very similar to diffusion. Thus in this particular model,
the effect of $\bar{\phi}$ on the velocity field is essentially to
provide some additional diffusion. Comparing with the last section, we
therefore expect that the Kolmogorov regime for the velocity mode
$k\bar{v}$ becomes smaller, since the effective diffusion becomes
stronger. 

In the inertial range eqs. \rf{sol} and \rf{ution} can be explicitly solved
analogously to what we did in the last section. Eq. \rf{ution} has
the solution
\beq
\bar{v}(k^2t)=R(27c\bar{v}(0)k^2t)/ck^2t,
\label{-3}
\eeq
where $R$ is given by eq. \rf{15}. Eq. \rf{sol} then gives
\beq
\bar{\phi}(k^2t)=\bar{\phi}(0)/\sqrt{1+4R(27c\bar{v}(0)k^2t)/c}.
\label{-6}
\eeq
Like in the last section it is easy to see that for large $k^2t$ eqs.
\rf{-3} and \rf{-6} imply the Kolmogorov and Obukhov-Corrsin exponents
for the velocity and for the field $\bar{\phi}=\psi$, respectively.
 
The advantage of the basic equations \rf{phidiffusion} and \rf{vdiffusion} is
that they are just two ordinary coupled
differential equations. Therefore, they are much simpler than the 
usual large number of coupled cascade equations. However, it should be
remembered that the set-up is very special. We need to assume initial ($t$=0)
spectra which are linear in $k$. Thus, it is not possible to start
e.g. with an initial Gaussian spectrum. However, the results in the discrete
cascade model reported in section 3 indicate that the final results
are independent of the initial state. However, in any case one may wonder
whether the simplicity has not been achieved at the cost of loosing
the physics of the problem. This will be discussed in the next section.

\section{Results in the continuous model}

In this section we shall compare the continuous cascade model with the
discrete one by obtaining qualitative and quantitative results. From the
scaling which we introduced for $v$ and $\phi$, it is clear that the
continuous model will produce an inverse cascade, whereby energy
is transferred from large to small $k$-values. Such a phenomenon is
known in magnetohydrodynamics (MHD) for the magnetic energy \cite{pouquet}
and in an MHD discrete cascade model \cite
{brandenburgetal}, where it is presumably due to an inverse cascade in
the three-dimensional magnetic helicity (for another shell-model
on this point, see \cite{Biskamp}). The continuous cascade model for
MHD \cite{b2} gives results very similar to the discrete model. In this
connection, it is of interest that the basic equations used for turbulent
mixtures are rather analogous to the MHD-equations.
  
We begin by a qualitative discussion of the results expected, based on
eqs. \rf{sol} and \rf{ution}. If diffusion is ignored (and $\alpha=0$)
these equations are explicitly solvable. Eq. \rf{-3} implies that 
$\bar{v}$ is given exactly like the velocity $u$ in section 5, so 
\beq
x\bar{v}(x)\propto x^{1/3}
\label{137}
\eeq
for large values of $x=k^2t$ in a range, where diffusion can still be
ignored. From eq. \rf{-6} we then see that the quantity
$C(k)=\psi^2/k=\bar{\phi}^2/k$ behaves like $k^{-5/3}$ with the
Obukhov-Corrsin exponent (with some time dependence). If the Prandtl number
$Pr=\nu/D$ is large, diffusion in the first place acts on the velocity.
Thus, $x\bar{v}$ starts to decrease, and from \rf {-6} it follows that
$\bar{\phi}$ approaches its initial value, i. e. $C(k)$ goes like 1/$k$
(with some time dependence), which is the Batchelor behaviour. Therefore,
without doing any numerical calculations, we see from eqs. \rf {ution} and
\rf{-6} that $C(k)$ must change slope from -5/3 to -1. When $k$ becomes so
large that the diffusion governed by $D$ is operative, of course $C(k)$ decays
exponentially, according to eq. \rf{sol}. If, on the other hand, $Pr$ is
of order one or less, it follows by the same reasoning as given above
(based on eqs. \rf{sol} and \rf{ution}) that only the -5/3 slope
materializes itself, before the exponential decays set in. This is
in agreement with the results obtained in the discrete model, as
discussed in section 3.

\begin{figure}[htb]
\vspace*{-3cm}
\mbox{{\epsfxsize=3.2in \epsfysize=5in \epsfbox{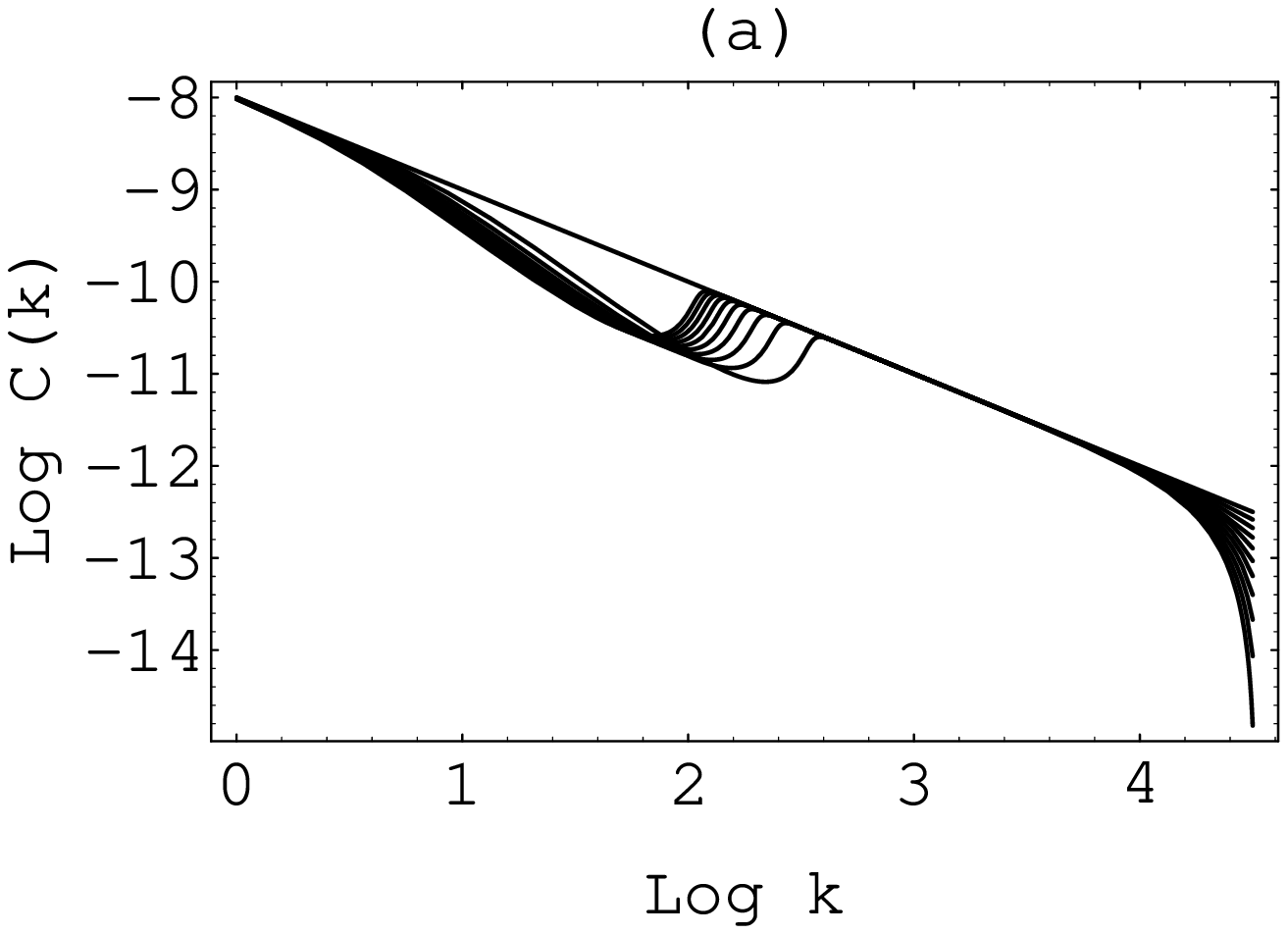}}
\vspace*{.5cm}
\hspace{.1cm}
{\epsfxsize=3.2in \epsfysize=5in \epsfbox{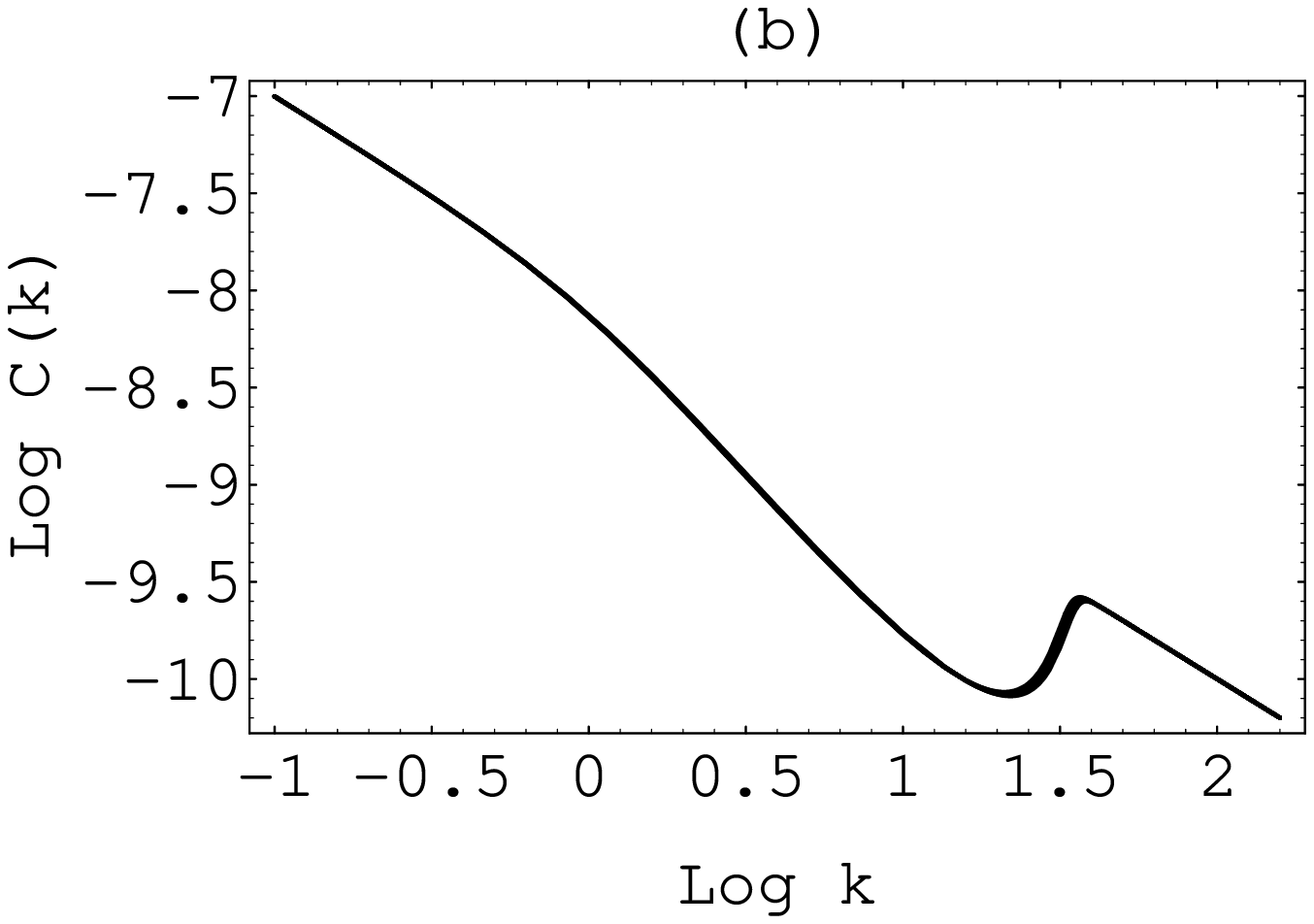}}}
\vspace*{-4cm}
\caption[yyy]{A double logarithmic (base 10) plot of the spectrum $C(k)$
as a function of $k$ in the continuous model for (a) ten small values
of time, from $t$=0 (where the $C$ goes like 1/$k$) to $t$=10$^{-2}$,
and (b) for hundred times larger. Time moves from the right to the left.}
\vspace*{0cm}
\hspace*{3cm}
\label{7_8}
\end{figure}

\begin{figure}[htb]
\vspace*{-3cm}
\mbox{{\epsfxsize=3.2in \epsfysize=5in \epsfbox{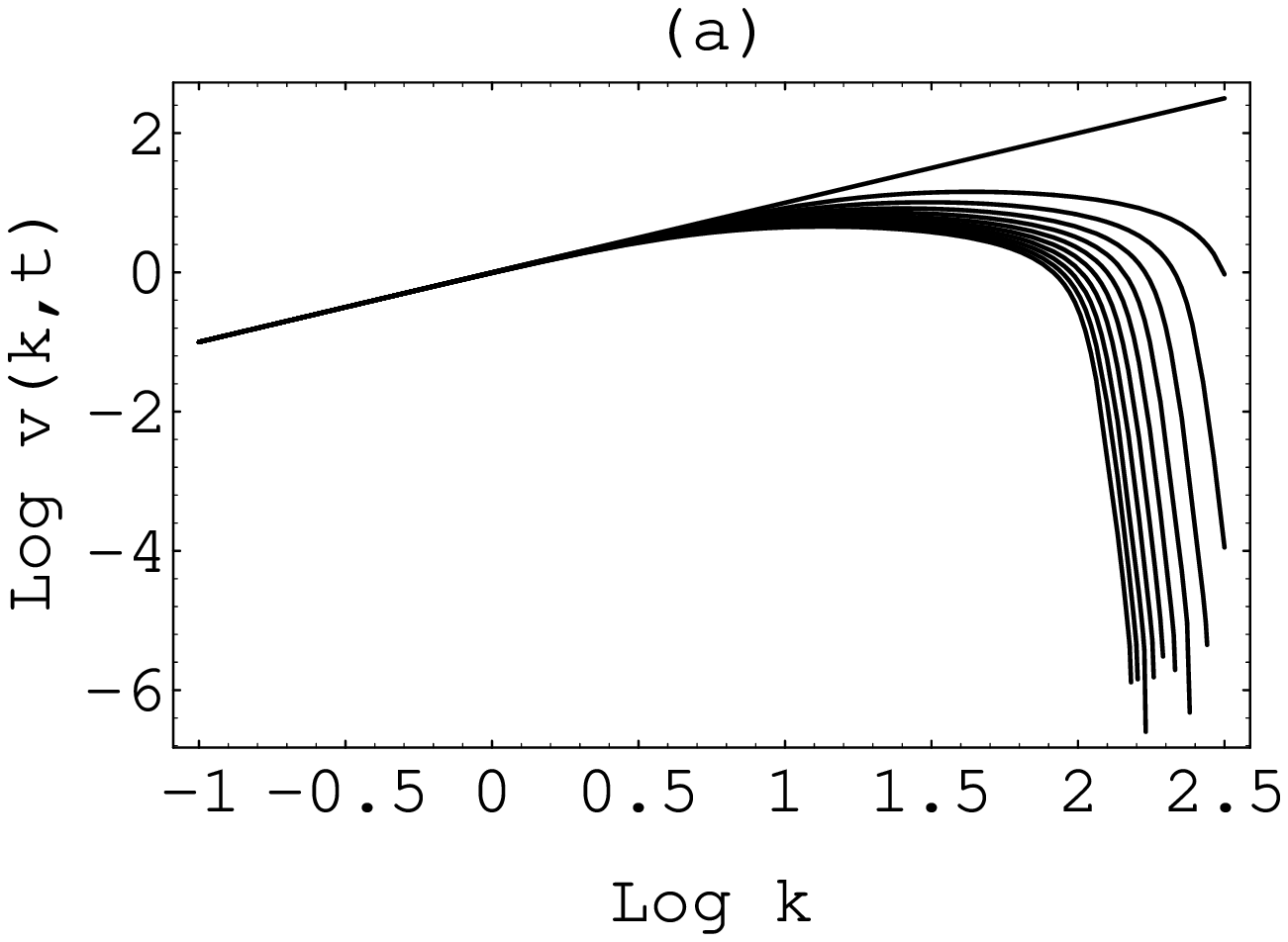}}
\vspace*{.5cm}
\hspace*{.1cm}
{\epsfxsize=3.2in \epsfysize=5in \epsfbox{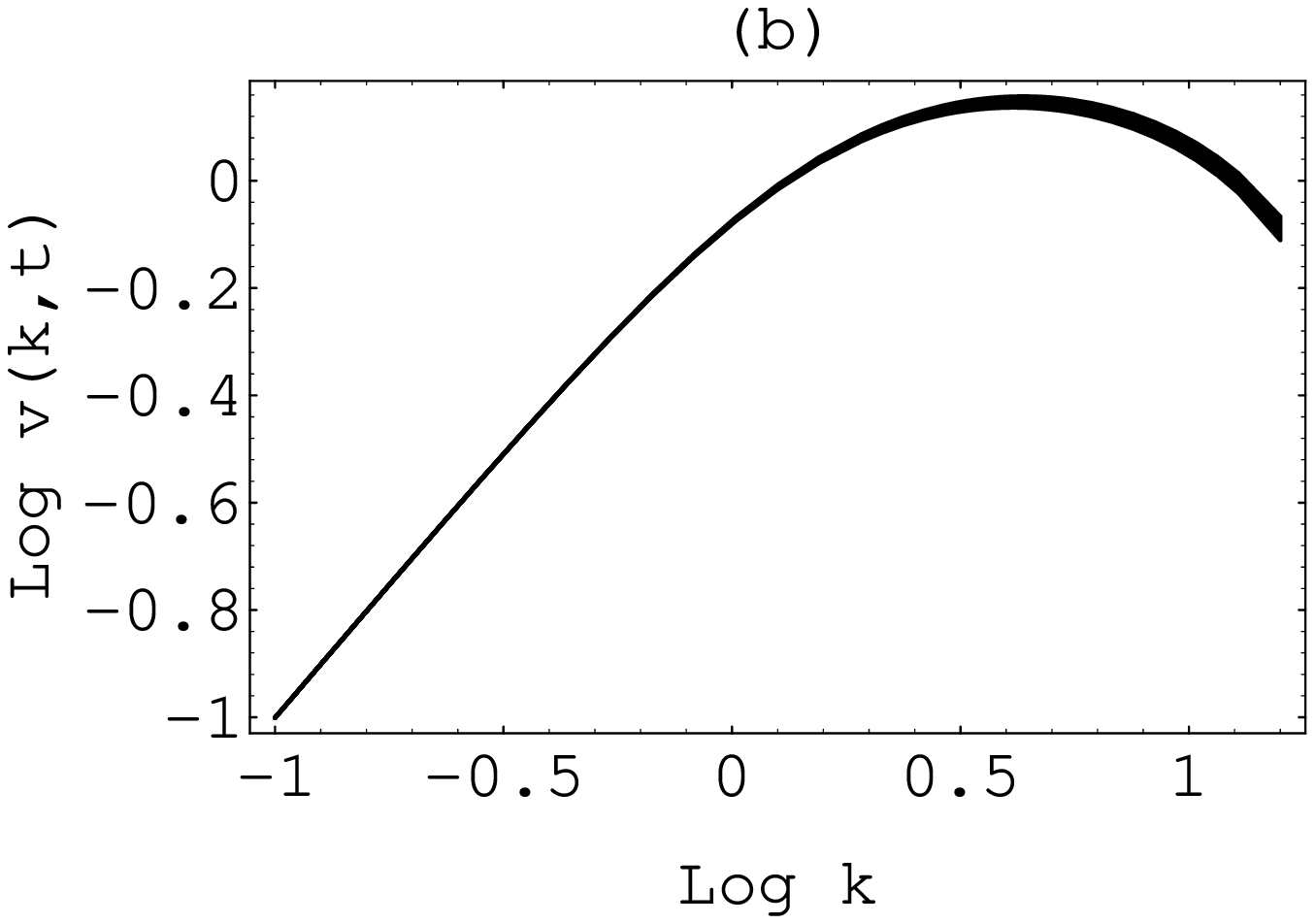}}}
\vspace*{-4cm}
\caption[yyy]{Double logarithmic plot (base 10) for the velocity
mode $v(k,t)=k\bar{v}$ as function of $k$ for (a)  ten time values between 0
(where $v(k,0)=k$) and 10$^{-2}$, and (b) for 100 times larger times. }
\vspace{0cm}
\hspace{3cm}
\label{7_9}
\end{figure}

We have made some numerical calculations, using the values
\beq
\nu={\rm 10}^{-1},~D={\rm 10}^{-7},~\alpha=\nu^2,~{\rm and}~Pr={\rm 10}^6.
\label{138}
\eeq
The initial values are given by $\bar{v}(0)$=1 and $\bar{\phi}$(0)=.0001. We
also took the constant $c$ in eq. \rf{vdiffusion} to be 1/2.
The results are presented for the spectrum $C(k)=\bar{\phi}^2/k=\psi^2/k$
and the velocity mode $k\bar{v}$. In Fig.\ \ref{7_8} we show $C(k)$ for
relatively low and for relatively large times. The initial behaviour is
linear in $k$ (from our initial condition that $\bar{\phi}$(0) is a
constant). From Fig.\ \ref{7_8} we see this behaviour for $t$=0. When $t$
increases, the slope changes to -5/3 in a range of $k$-values from
approximately 10 to slightly less than 10$^2$. From Fig.~\ref{7_9} for 
the velocity mode, we see that at a $k$-value around 100,
the velocity decreases rapidly. Therefore, as one can see from Eq. \rf{sol}
and as one can also see from Fig.\ \ref{7_8}(a),
after the initial decrease with slope -5/3, the spectrum $C(k)$ increases
until it regains the Batchelor slope -1. Ultimately, for $k$ slightly above
10$^4$ one sees that $C(k)$ starts to decrease exponentially. This repeats
itself at later times, as is also seen from Fig.~\ref{7_8}, but when time
increases, the ``velocity of change'' $k/t\propto 1/\sqrt{t}$ decreases,
so the curves for the ten different times in Fig.~\ref{7_8} are much closer
at later times. In the region where $C$ increases (between slope -5/3 and
slope -1), there are very short range correlations in $\bf x$-space. This
could be a rudimentary version of an intermittency fluctuation.

We have also investigated
the question of equipartition. For small $k$ the kinetic energy dominates.
Around the value of $k$ where the peak appears, there is equipartition in
essentially only one point. The energy $E_{\psi}=\alpha k\psi^2$ then dominates
for larger values, where the kinetic energy becomes very small compared to
$E_{\psi}$. It should also be noticed that the peak is time dependent,
and moves towards smaller values of $k$ with a ``velocity'' $k/t=1/\sqrt{t}$.

Finally, we have studied the influence of the parameter $\alpha$. It seems
that the continuous model differ from the discrete one with respect to
this point. For the various values of $Pr$ and $\nu$ we have
studied, we find that the influence of $\alpha$ is only rather marginal, and
we do not find any really spectacular effect of $\alpha$. It
would therefore be interesting to study the case where the initial velocity
is zero (or very small). In this case there is no effect (or only a slight
effect) if we have a passive scalar, so with $\alpha$ non-vanishing one will
see an effect (almost) entirely due to $\alpha$. This problem can be analyzed
from eqs. \rf{sol} and \rf{ution} by replacing $\bar{v}$(0) by
$\bar{v}(\epsilon)$ and performing the limit $\epsilon\rightarrow$0 in such a
way that $\bar{v}$ approaches zero. We can also study this by analyzing
eq. \rf{vdiffusion} near $x$=0, using the boundary condition $\bar{v}$(0)=0.
We get from \rf{vdiffusion} (compare with eq. \rf{mny})
\beq
\bar{v}(x)\approx -2\alpha \bar{\phi}(0)^2~x+O(x^2),
\label{137137}
\eeq
Thus the velocity scaling function must be negative and linear for small 
values of the scaling variable $x=k^2t$. 

\begin{figure}[htb]
\vspace*{-3cm}
\centerline{\epsfxsize=3.5in \epsfysize=5in \epsfbox{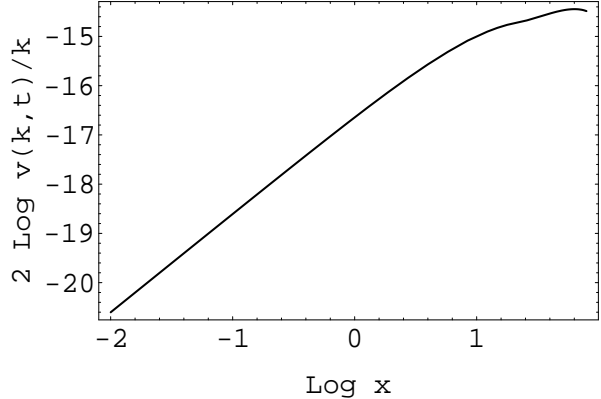}}
\vspace*{-3cm}
\caption[iii]{A log-log plot (base 10) of the scaling function
$\bar{v}^2=(v(k,t)/k)^2$ as a
function of the scaling variable $x=k^2t$ for the initial conditions
$\bar{v}(0)$=0 and $\bar{\phi}(0)=\psi(0)$=.0005.}
\label{7_58}
\end{figure}

We have also studied this problem numerically. In Fig.\ \ref{7_58} we show the
resulting scaling function $\bar{v}^2=v(k,t)^2/k^2$ as a function of the
scaling variable. The values of $\nu$ and $Pr$ are as used before, but
initially we assume $\bar{v}$(0)=0, and $\bar{\phi}(0)=\psi(0)$=.0005.
Although the velocity-scaling function is very small, there is clearly an
effect. However, the resulting back-reaction on $\bar{\phi}$ is small
in most cases. This can be seen from eq. \rf{sol}, since the velocity field in
the denominator on the right hand side is small. However, there may
exist initial conditions where this is not true, and where $x\bar{v}(x)$
may approach -1, causing $\psi$ to diverge. In such a case stabilizing terms
of higher orders in $\psi$ must be included in the basic eqs. \rf{m5} and
\rf{m6}. Disregarding this possibility, the density
function $\psi$ is rather insensitive to velocity fluctuations, when the
kinematic viscosity $\nu$ is much larger than the viscosity $D$ of $\psi$.
This effect is somewhat similar to what has recently been seen in
MHD \cite{b2,Fauve} for large ``Prandtl'' numbers 
(i.e. for the kinematic viscosity much larger than the Ohmic diffusion). 

{}From an experimental point of view it would be interesting to study the case
where the initial velocity vanishes or is small, and where some initial
distribution (e.g. random) is established with a non-constant  gradient of
$\psi$. The fluids should then be set ``spontaneously'' in motion. The
gradient field is analogous to the magnetic field in MHD. Therefore, such an
experimental set up would be somewhat analogous to the study of primordial
magnetic fields in the early universe (see refs. \cite{brandenburgetal} and
\cite{b2}), where the magnetic field can induce a velocity field.
It would clearly be of interest to investigate this analogy in an earthbound
laboratory.

To conclude this section, we do not find the large peak predicted in
ref. \cite{Ruiz2} for increasing Prandtl numbers. There is a small
time dependent peak of magnitude less than a half decade,
as can be seen e.g. from Fig.~ \ref{7_8}. This is to
be contrasted with a peak of several decades in ref.
\cite{Ruiz2}. Whether this is a shortcoming of the continuous model
is, of course, an experimental question. However, our results are not so
different from those of the discrete GOY model, discussed in section 3,
where the peak is of the same order of magnitude as found here. However,
the nature of the peak is different in the two cases, since the $\alpha$
dependence differ, as mentioned above. 

\section{Conclusions}

The main result of the present paper is that intermittency effects 
are likely to play an important role 
in turbulent binary fluids. The influence of intermittency is not small;
compared to previous studies of non-intermittent binary fluids
\cite{Ruiz1,Ruiz2} the peak in the concentration spectrum is much less
pronounced and less persistent because the fluctuations
tend to ``surround'' and diminish the peak. It is therefore not easy to see
the effect of the active coupling term \rf{m6} from the spectra, in accordance
with experiments \cite{Walter2}-\cite{Walter4}. The transport
coefficient $\alpha$ in \rf{m6} presumably has its most dramatic effect
in the case where the initial velocity vanishes. Here the existence of
a non-vanishing $\alpha$ implies the ``spontaneous'' generation of a velocity
field, provided there is an initial variation in the gradient of $\psi$.
The observation of such an effect would have an analogy in MHD, where an
initial (``primordial'') magnetic field induces a velocity field, which may
be of relevance in the early universe. On the other hand, if such an effect
is not observed in binary mixtures, this would indicate that the active
coupling term \rf{m6} is probably not present. If so, binary mixtures would
not be analogous to MHD.

Also, we see that the continuous model gives results which are rather similar
to the discrete model. In the continuous case there also
exists a time-dependent peak. 

\vskip0.4truecm

We thank David R. Nelson and Walter Goldburg for stimulating discussions. 
We also thank George Savvidy
for reminding us of Heisenberg's paper \cite{heisenberg}, and Yuri Makeenko
for much help in inserting the figures.

\end{document}